\documentclass[sigconf,9pt]{acmart}

\AtBeginDocument{%
  \providecommand\BibTeX{{%
    \normalfont B\kern-0.5em{\scshape i\kern-0.25em b}\kern-0.8em\TeX}}}

\setcopyright{acmcopyright}
\copyrightyear{2021}
\acmYear{2021}
\acmDOI{10.1145/XXXXXXX.XXXXXXX}

\acmConference[e-Energy'21]{ACM e-Energy 2021}{June 28--July 2, 2021}{Torino, Italy}
\acmBooktitle{Proceedings of ACM e-Energy Conference,
  June 28--July 2, 2021, Torino, Italy}
\acmPrice{15.00}
\acmISBN{978-1-4503-XXXX-X/YY/MM}

\usepackage{amsmath,bm}
\usepackage{graphicx,subcaption}
\usepackage[ruled,vlined]{algorithm2e}
\usepackage{algorithmic}
\usepackage{booktabs}
\usepackage{comment}

\theoremstyle{plain}
\newtheorem{theorem}{\protect\theoremname}
\theoremstyle{plain}
\newtheorem{definition}[theorem]{\protect\definitionname}
\theoremstyle{plain}
\newtheorem{proposition}[theorem]{\protect\propname}
\theoremstyle{plain}
\newtheorem{lemma}[theorem]{\protect\lemmaname}
\providecommand{\definitionname}{Definition}
\providecommand{\propname}{Proposition}
\providecommand{\lemmaname}{Lemma}
\providecommand{\theoremname}{Theorem}

\ifodd 0
    \newcommand\rev[1]{{\color{red}#1}}
    \newcommand{\com}[2]{\textbf{\color{blue} (COMMENT from [#1]: #2)}}
\else
    \newcommand\rev[1]{{#1}}
    
    \newcommand{\com}[2]{}
\fi

\def\proname{{\textsf{PRM}}}
\def\algname{{\textsf{pCR-PRM}}}

\begin{document}

\title{Optimal Online Algorithms for Peak-Demand Reduction Maximization with Energy Storage}

\orcid{1234-5678-9012}
\author{Yanfang Mo, Qiulin Lin, Minghua, Chen, and Si-Zhao Joe Qin}
\affiliation{%
  \institution{School of Data Science}
  \institution{City University of Hong Kong}
  %\streetaddress{City University of Hong Kong}
  \city{Kowloon}
  \state{Hong Kong}
  \country{China}
  \postcode{999077}
}
\email{{yanfang.mo, qiulin.lin, minghua.chen, joe.qin}@cityu.edu.hk}

\renewcommand{\shortauthors}{Mo et al.}

\begin{abstract}

The high proportions of demand charges in electric bills motivate large-power customers to leverage energy storage for reducing the peak procurement from the outer grid. Given limited energy storage, we expect to maximize the peak-demand reduction in an online fashion, challenged by the highly uncertain demands and renewable injections, the non-cumulative nature of peak consumption, and the coupling of online decisions. In this paper, we propose an optimal online algorithm that achieves the best competitive ratio, following the idea of maintaining a constant ratio between the online and the optimal offline peak-reduction performance. We further show that the optimal competitive ratio can be computed by solving a linear number of linear-fractional programs. Moreover, we extend the algorithm to adaptively maintain the best competitive ratio given the revealed inputs and actions at each decision-making round. The adaptive algorithm retains the optimal worst-case guarantee and attains improved average-case performance. We evaluate our proposed algorithms using real-world traces and show that they obtain up to $81\%$ peak reduction of the optimal offline benchmark. Additionally, the adaptive algorithm achieves at least $20\%$ more peak reduction against baseline alternatives.

\end{abstract}


\begin{CCSXML}
<ccs2012>
   <concept>
       <concept_id>10010147.10010178.10010199.10010201</concept_id>
       <concept_desc>Computing methodologies~Planning under uncertainty</concept_desc>
       <concept_significance>500</concept_significance>
       </concept>
   <concept>
       <concept_id>10003752.10003809.10010047</concept_id>
       <concept_desc>Theory of computation~Online algorithms</concept_desc>
       <concept_significance>500</concept_significance>
       </concept>
   <concept>
       <concept_id>10010405.10010481.10010484</concept_id>
       <concept_desc>Applied computing~Decision analysis</concept_desc>
       <concept_significance>500</concept_significance>
       </concept>
 </ccs2012>
\end{CCSXML}

\ccsdesc[500]{Computing methodologies~Planning under uncertainty}
\ccsdesc[500]{Theory of computation~Online algorithms}
\ccsdesc[500]{Applied computing~Decision analysis}

\keywords{energy storage management, peak-demand charge, online competitive algorithms}


\maketitle

%%%%%%%%%%%%%%%%%%%%%%%%%%%%%%%%%%%%%%%%%%%%%%%%%%%%%%%%%%%%%%%%%%%%%%%%%%%%%%%%%%%%%%%%%%%%%
\section{Introduction}
\begin{figure}[t]
  \centering
  %\centerline{\includegraphics[width=0.86\columnwidth, height=0.32\columnwidth]{figures/ExampLoadScenario.pdf}}%
  \centerline{\includegraphics[width=0.9\columnwidth]{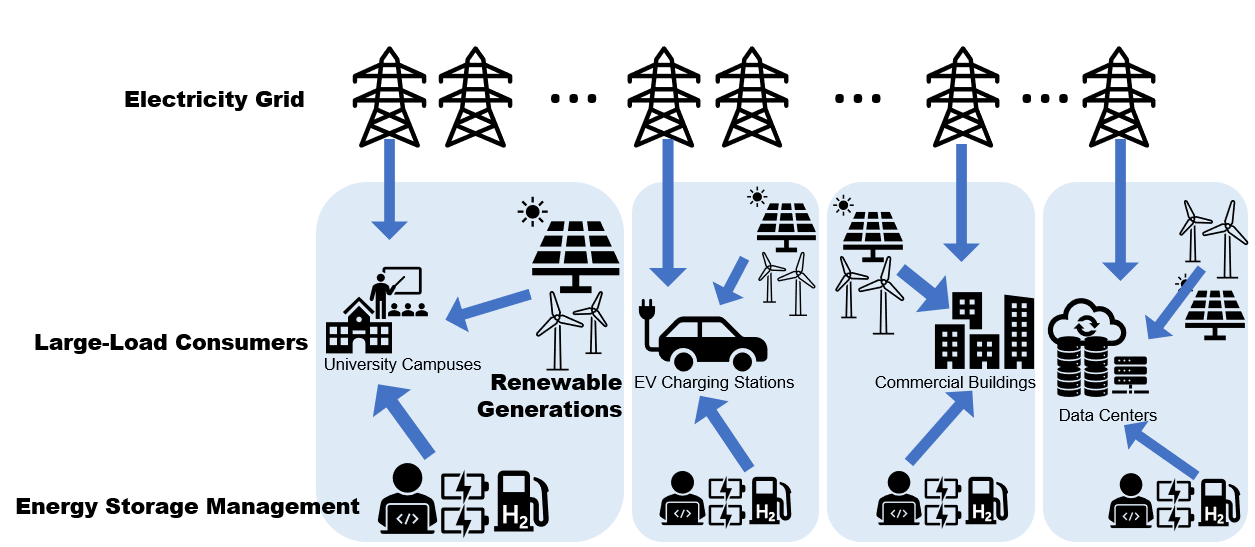}}%
  \caption{Illustrative scenarios where large-load consumers own renewable generations and energy storage systems. }\label{fig: ExampLoad}
\end{figure}

Electricity bills have been a heavy burden on industrial or commercial consumers~\citep{sun2010demand}. Typically, commercial buildings spend~$1.44$ dollar per square foot per year on electricity~(matching the management fee) according to the Commercial Building Energy Consumption Survey~\citep{Average_Utility_Cost}, while the electricity consumption accounts for up to~$70\%$ of the total operating costs of a data center~\citep{Power_Data_Center, Power_Data_Center2}. The rapid growth of electricity expenses usually relates to higher greenhouse air emissions and goes against sustainable development. Owing to the financial and environmental concerns, these large-load customers have strong desires to study the evolutionary pricing strategies and reduce electricity costs accordingly.

As the introduction of demand side management in the~$1980$s~\citep{gellings1985concept}, utilities tend to exploit delicate pricing schemes for motivating customers to modify their power consumption patterns. As a result of the evolution, a large-power consumer's electricity bill generally consists of two parts: volume charge and demand charge~\citep{Ex_bill}. The volume charge is mainly due to the amount of consumed energy. Moreover, utilities usually adopt time-based rates like Time-of-Use~(TOU) pricing~\citep{newsham2010effect} to reflect the energy costs over time. \rev{However, such pricing strategies may cause rebound peaks, which calls for the second part of the bill~\cite{chakrabortty2012control}.} % the higher unit prices are not sufficient to incentivize consumers to reduce their electricity usage during peak hours. Meanwhile, the augmented demand peak often incurs significant costs (e.g., investing in peaking power plants) and damages (e.g., transmission line overloading) to power systems~\citep{oren1992design,chakrabortty2012control}. These facts necessitate the second part of the bill.

The peak-demand charge refers to the extra punitive charge on the maximum amount of power drawn for any prespecified interval, e.g.,~$15$ or~$30$ minutes, during a billing period~\citep{Ex_bill}. The introduction of peak-demand charge motivates large-power customers to flatten their demand curves. Unlike volume charges~($\text{kWh}\times \$/\text{kWh}$), the peak-demand charge rate is in dollar per kilowatt. According to~\cite{CLP}, the demand charge rate and on-peak volume charge rate are around~$118$ HK dollars/kW and~$55$ HK cents/kWh respectively. It follows that the cost of lifting the peak demand is roughly~$200$ times that of increasing the off-peak energy consumption by one unit. Moreover, we see that the peak-demand often takes up a large portion of the bill, e.g., up to~$90\%$ for DC fast charging stations~\citep{lee2020pricing} and~$80\%$ for {G}oogle data centers~\citep{xu2014reducing}. Therefore, we should pay particular attention to the peak-demand charge.

Two direct approaches to electricity cost saving are reducing peak demands and shifting partial loads from an on-peak period to an off-peak one~\citep{liu2013data,xu2014reducing,shi2017using}. However, such methods are not always applicable, as certain loads cannot be cut or shifted. Meanwhile, bulk energy consumers increasingly invest in self-owned renewable generations, e.g., solar power systems, for green buildings and industries. These installations can reduce the amount of energy purchased from the grid. However, they may result in a more fluctuating net demand curve and not reduce the peak-demand charge because the renewable generations are highly volatile~\citep{jacobsen2012curtailment}.

On the other hand, the rapid development of storage technologies makes it convenient and economical to build and maintain a storage system. Energy storage has also been useful to meet demand surges and provide uninterrupted power supply in power systems~\citep{sobieski1985economic,urgaonkar2011optimal,guo2013electricity}. These facts inspire us to discharge the stored energy in fuel cells or batteries to reduce the peak-demand charge.

As illustrated in Fig.~\ref{fig: ExampLoad}, this work targets large-load customers with self-owned renewable generations, and our objective is to maximize the peak-demand reduction by using energy storage in an on-peak period. \rev{First note that the volume charge prices are much lower in off-peak periods, so we had  better fully charge the storage system then. Second, the on-peak periods of neighbour users often coincide. Thus, recharging may increase the cumulative peak demand in a local grid, which is not socially friendly. Third, energy storage systems, like pumped storage, compressed-air storage, and fuel cells, can not be recharged during use. These facts motivates us to focus on discharging in on-peak periods. Last but not least, our study on storage discharging can shed light on the joint charging/discharging optimization case.}% We focus on discharging the energy storage  for the following reasons.

If we know the future demands, it is easier to attain the best use of energy storage. Unfortunately, the small-scale demands are highly uncertain, and self-owned renewable generations even exacerbate the volatility of demands over time~\citep{jacobsen2012curtailment}. Consequently, it is challenging to schedule stored energy in real-time to flatten the net demand curves. Furthermore, the non-cumulative nature of peak consumption increases the difficulty in maximizing the peak-demand reduction by using energy storage in an online fashion.

The unpredictability of net demands prevents us from managing stored energy by stochastic optimization or model predictive control. The resulting algorithms may overfit to the estimated model and be sensitive to a volatile practical environment~\citep{urgaonkar2011optimal,ma2012demand,alharbi2017stochastic,qin2015online,risbeck2020economic}. Contrary to these, robust optimization relies little on accurate estimations. It models uncertainties by a set and emphasizes the worst-case outcome of proposed algorithms under the uncertainty set~\citep{bertsimas2011theory,xiang2015robust}. However, such an approach may suffer from heavy computation and be too conservative.

In our paper, we apply another popular and useful tool, online algorithm design with competitive analysis~\citep{borodin2005online,buchbinder2016unified}. In an online setting, an input sequence will be revealed sequentially in time, and the decision maker has to make irrevocable decisions at current time with little or no future information.  In our problem, an input sequence refers to a series of demands over time. While stochastic optimization and robust optimization emphasize the absolute performance, competitive online optimization concerns the relative performance compared to the optimal offline outcomes. Specifically, it evaluates an online algorithm via Competitive Ratio (CR), which relates to the worst-case ratio between the offline and online outcomes under the same input sequence. In other words, competitive online optimization considers the robustness by conducting the worst-case analysis and \rev{fairness in performance evaluation under different input sequences by adopting the offline-to-online relative performance.} More importantly, the best possible CR captures the price of uncertainty for the considered online problem. These merits also indicate the difficulty of applying competitive analysis. To identify the worst-case relative performance and develop online algorithms with the best CR, we frequently encounter nonconvex and combinatorial problems at first sight. Despite these challenges, we shall apply competitive online optimization to the storage-assisted peak-reduction maximization problem under our consideration. We make the following contributions.%In the following, we outline the rest of the paper and our main contributions.

%$\mbox{\ensuremath{\rhd}}$ Section~\ref{Sec_RelatedWork} is dedicated to related work. We defer the proofs to the Appendix for fluency.

$\mbox{\ensuremath{\rhd}}$ In Section~\ref{Sec_ProbForm}, we rigorously define the problem as finding an energy storage utilization scheme that maximizes the peak-demand reduction. %\footnote{\rev{The consideration of maximizing the peak-demand  reduction differs from that of peak-demand minimization in the following aspects. First, it highlights the benefit introduced by energy storage. Second, it removes the impact of the peak-demand before using the storage making it comparable among different demand profiles. Third, in the online competitive analysis that we apply, cost minimization and benefit maximization are different\cite{Ming2019online,yi2019balancing}.}}.
Moreover, we uncover the inherent structural insights into the optimal offline solution.

$\mbox{\ensuremath{\rhd}}$ In Section~\ref{Sec_OnlineAlg}, we design the first and optimal online algorithm that achieves the best CR. Our algorithm chooses actions to maintain a constant ratio between the online and offline performance at each round. We further show that the best feasible ratio to maintain is also the optimal CR of the problem. %In this way, the algorithm guarantees the optimal worst-case performance relative to the offline optimal counterpart.
Notably, we propose an efficient numerical approach to computing the best CR by a linear number of linear-fractional programs. %Moreover, these programs are not dependent on each other, and we can solve them in parallel.

$\mbox{\ensuremath{\rhd}}$ In Section~\ref{Sec_Adaptive}, we extend our approach to design an adaptive online algorithm that achieves adaptively-improved average-case performance while retaining the worst-case performance guarantee. At each round, the revealed input prunes the uncertainty set and makes it possible to obtain a better CR in the remaining rounds. The adaptive algorithm utilizes the idea and maintains the best possible CR at each round given the revealed input and online actions so far. The adaptive algorithm is among the few online approaches that respond to the non-occurrence of worst-case input and adaptively improve the average-case performance. %This practice combines efficiency and robustness, so the adaptive extension can improve the average-case performance and retain the optimal worst-case performance. Recall that the competitive analysis focuses on the worst-case relative performance, and the best CR captures the price of uncertainty before we know any inputs. Few competitive online algorithms respond to the nonoccurrence of worst cases, while the price of future uncertainty may change as we observe more data. In contrast, our adaptive algorithm actively exploits the real-time information and pursues the instantaneous prices of future uncertainty slot by slot.%an improved offline-to-online ratio slot by slot.  for a better average-case performance, and still guarantee the optimal worst-case performance. Specifically, the adaptive algorithm captures the prices of future uncertainty over time and pursues improved offline-to-online ratios accordingly. In this way, we can combine efficiency with robustness in practice.

$\mbox{\ensuremath{\rhd}}$ In Section~\ref{Sec_Simulation}, we conduct simulations on real-world traces and evaluate the empirical performance of our algorithms in representative scenarios. Simulation results show that our online algorithms improve the peak-demand reduction by more than $20\%$ as compared to baseline alternatives. Further, with little future information, they achieve up to $81\%$ of the (best possible) peak reduction attained by the optimal offline algorithm.

\rev{Due to page limitations, omitted proofs can be found in our technical report.}

%%

%%%%%%%%%%%%%%%%%%%%%%%%%%%%%%%%%%%%%%%%%%%%%%%%%%%%%%%%%%%%%%%%%%%%%%%%%%%%%%%%%%%%%%%%%%%%%
\section{Related Work}\label{Sec_RelatedWork}
We discuss  three lines of related works in the following.
\paragraph{Using energy storage}
As mentioned in the Introduction, \rev{more} customers exploit energy storage for cost-savings on their electric bills and possible arbitrage opportunities. This phenomenon raises researchers' interests along this line, as exemplified below. \cite{walawalkar2007economics} assessed the storage value for energy arbitrage and regulation services in New York. \cite{urgaonkar2011optimal} explored time-average cost reduction opportunities of using existing storage, like UPS units of data centers. Another kind of existing storage refers to electric vehicles, and the economics of vehicle-to-grid services has been examined in~\cite{white2011using}. The storage is valuable not only for commercial consumers, but also for residential customers, as discussed in~\cite{leadbetter2012battery,ma2012demand} and~\cite{xi2014stochastic}. In addition to cost-saving, \cite{Rishikesh2020Emission} investigated reducing emission footprint of the power grid using distributed energy storage. In short, researchers are exploring the applications and potential financial interests of storage from various aspects.

\paragraph{Peak-demand charge}
The peak-demand charge has attracted attention from both utilities and large-load customers. \cite{neufeld1987price} analyzes the adoption of demand charge given the competition of isolated industrial customers. \cite{zakeri2014optimization} considered an extended peak-demand charge which relates to the largest accumulated consumption over several periods. In the literature, researchers have studied multiple approaches for utilities and large-load customers in response to the peak-demand charge. Based on day-ahead load predictions, \cite{shi2017using} studies the joint optimization of peak shaving and frequency regulation with energy storage.  Under a peak-based pricing model, \cite{zhang2016peak} considers the online economic dispatching of local generators in microgrids. \cite{zhao2013peak} and~\cite{lee2020pricing} respectively examine the scheduling and the pricing of electric vehicle services under the electricity tariff with peak-demand charge. See a survey in \cite{UDDIN2018review} for more examples. Overall, the peak-demand charge brings utilities and consumers together for efficient and reliable power systems. In our paper, we consider that a large-load consumer utilizes his or her energy storage to reduce the peak-demand charge. We focus on maximizing the peak-demand reduction brought by using energy storage. Compared with peak-demand minimization that considers the absolute value, the reduction reflects the relative value between the peak-demands before and after using energy storage. It allows us to make a fair comparison on the benefit of the storage under different demand profiles.

\paragraph{Online competitive analysis}
Online competitive analysis has been useful in electric vehicle charging~\citep{zhao2013peak,tang2014online,yi2019balancing} and economic dispatching~\citep{zhang2016peak,chau2016cost}. The competitive analyses of cost minimization and benefit maximization have different challenges, as discussed in \cite{Ming2019online} and \cite{yi2019balancing}. Although we can easily transform the problem from one to the other in the offline scenario, solutions and results for one problem may not directly apply to its counterpart. %An online algorithm achieving strong performance guarantee in one case
In this paper, we consider online peak-demand reduction maximization. %which focus on the relative value of the peak-demands before and after using the storage instead of the absolute peak-demand in peak-demand minimization. %The competitive ratio in such case,  depending on the size but not the absolution value  of the demand range, directly reflects the price of uncertainty.
%The online competitive analysis in our problem also demonstrates differences with the peak-demand minimization.
%\footnote{We will discuss the difference in more detail after we formulate our problem.}

We adopt a similar algorithmic framework to that of~\citep{yi2019balancing} and~\citep{lin2019competitive} for the proposed online storage management problem. The resulting algorithms are parameterized by a ratio ``pursued'' in each decision-making round, and we can actively adapt the ratio for real-time information. It turns out that our algorithms are effective and efficient from theoretical analysis and empirical validations.

%%
%%%%%%%%%%%%%%%%%%%%%%%%%%%%%%%
\section{Problem Formulation}\label{Sec_ProbForm}
In this section, we elaborate on the model of applying energy storage to reduce the peak-demand. We formulate the problem as a peak-demand reduction maximization problem under a capacity constraint and discharging rate constraint. We also introduce the offline optimal solution to the problem given the input is known in advance and \rev{emphasize} our focus on the practical online scenario. In Table~\ref{table: acronyms}, we summarize the useful symbols and terms for future reference. %For clarity, we summarize in  the symbols and terms used in this paper. For fluency, we defer proofs to the appendix.% and characterize its optimal offline solution.
\begin{table}[t!]
  \centering
  \caption{A summary of symbols.}\label{table: acronyms}
  \begin{tabular}{p{0.15\columnwidth}p{0.75\columnwidth}}
    \toprule %\hline
    Symbols & Meanings \\ \midrule %\hline
    %$\bm 1$ & an all-one vector with a compatible dimension\\ \hline
%    $[x]^+$ & $\max\{x,0\}$, for~$x\in \mathbb{R}$ \\ %\hline
   % $\lfloor x\rfloor$ & the largest integer no more than~$x$, for~$x\in\mathbb{R}$\\ %\hline
%    $[n]$ & $\{1,2,\ldots,n\}$, where~$n\in\mathbb{N}$ and~$[0]$ reduces to the empty set\\ %\hline
    %$|\mathcal{I}|$ &  cardinality of a set~$\mathcal{I}$ \\ \hline
    %$\mathcal{X}\backslash \mathcal{Y}$ &  $\{x\mid x\in \mathcal{X}~\&~ x\notin \mathcal{Y}\}$ \\% \hline
    $c$ & scaled storage energy capacity (kWh)\\% \hline
    $T$ & number of time slots\\ %\hline
    $\bm d$ & a demand profile,~$\bm d = (d_1,d_2,\cdots,d_T) \in \mathbb{R}^T$, where~$d_t$ is the net demand at time slot~$t$ (kWh)\\ %\hline
    %$d_t$ & net demand at time slot~$t$ (kWh)\\% \hline
    $\underline{d},\overline{d}$ & lower and upper bounds of the net demand at a time slot (kWh)\\ \hline
    $\delta_t$ & discharging quantity at time slot~$t$,~$\delta_t\geq 0$ (kWh) \\ %\hline
    $\bar{\delta}$ & maximum discharging quantity per time slot (kWh)\\ %\hline
    %$\eta$ & storage discharging efficiency, $0<\eta\leq 1$\\ \hline
    $v(\bm d)$ & peak usage of the offline optimal solution under demand profile~$\bm d$ \\ %\hline
    $\sigma(\bm d)$ & peak-demand reduction of the offline optimal solution under demand profile~$\bm d$ \\ %\hline
%    $\sigma^{\mathfrak{A}}(\bm d)$ & Peak-demand reduction under an online algorithm~$\mathfrak{A}$ and a demand profile~$\bm d$ \\% \hline
%    $\bm d^t$ & The reference input~$[d_1~d_2~\cdots~d_t~\underline{d}~\cdots~\underline{d}]'$ for time slot~$t$\\ %\hline
    $\bm \delta(\pi,\bm d)$ & solution of \algname($\pi$) under demand profile~$\bm d$\\ \hline
    $\pi^*$ & the best possible CR among all online algorithms \\ %\hline
    $\pi^*_t$ & adaptive CR at time slot~$t\in[T]$ \\ \hline
%    $\textsf{CR-Comp}(\mathcal{I})$ & An auxiliary linear-fractional program for~$\pi^*$\\% \hline%, indexed by~$\mathcal{I}$ \\ \hline
%    $\textsf{AdaCR-Thre}(\pi,\mathcal{I})$ & An auxiliary linear program for~$\pi^*_t$\\ \bottomrule%, indexed by~$(\pi,\mathcal{I})$\\ \hline
   \end{tabular}
\end{table}

%%%%%%%%%%%%%%%%%%%%%%%%%%%%%%%%%%%
\subsection{Mathematical Model}
\rev{We assume that the peak demand of a large-power customer may occur in~$T$ time slots of an on-peak period, like from~$9$ a.m. to~$10$ p.m. for a shopping mall~\cite{DayNightTariffs}. Each time slot corresponds to~$15$ or~$30$ minutes by the power measurement. Then, we can consider the consumed energy (in kWh) instead of the power demand (in kW).}

\paragraph*{Net electricity demand:} We consider a general setting where the excess renewable generation cannot recharge the storage system. The self-owned renewable generations are not sufficient to cover the gross demand of the large-load customer during the~$T$ slots. Then, we use~$\bm d\in \mathbb{R}^T$ to denote the net demand profile, which is given by the difference between the gross demand and the renewable generation. Although the cost of renewable generation is negligible, the volatile renewable generations exacerbate the unpredictability of net demands over time. Consequently, it is hard to estimate~$d_t$ accurately before time slot~$t$. We just know the lower and upper bounds of the net demand in a time slot. For brevity, we herein focus on the case with uniform bounds, namely~$d_t\in[\underline{d},\overline{d}]$ for all~$t\in[T]$,\footnote{In this paper, we use $[n]$ to denote $\{1,2,\cdots,n\}$ where~$n\in\mathbb{N}$ and~$[0]$ reduces to the empty set.} while our approach also applies to the case with time-variant bounds after slight modifications.

%%%%%%%%%%%%%%%%%%%%%%%%%%%%%
\paragraph*{Energy storage:}
% We discharge the stored energy (e.g., in fuel cells) to partially meet the net demand and reduce the electricity purchase from the grid in each time slot. The focus on discharging can simplify the storage model.
 Let~$c$~(in~kWh) be the energy storage capacity scaled by a factor concerning the maximum depth of discharge and the discharging efficiency. Let~$\bar{\delta}$~(in~kWh) denote the maximum discharging quantity per time slot. In practice, we can hybridize different energy storage technologies with complementary characteristics~\citep{hemmati2016emergence}. For example, flywheel and supercapacity are high power devices, while fuel cells and pumped hydro are high energy devices. In this way, we can attain desired~$c$ and~$\bar{\delta}$ to meet long-term energy needs and short-term power needs. We consider the scenario that we have fully charged the energy storage before the on-peak period and discharge the storage in the on-peak duration to reduce the peak-demand.  We denote the discharging vector by~$\bm \delta = (\delta_1,\delta_2,\cdots,\delta_T)\in \mathbb{R}^T$, where~$\delta_t$~(in~kWh) is the discharging quantity at time slot~$t$. Then, the characteristics of the storage system leads to the inventory constraint~$\sum_{t=1}^{T}\delta_t \leq c$ and the rate constraints~$\delta_t\leq \bar{\delta}$, for all~$t\in [T]$.

\subsection{Problem Formulation}\label{sec:profor}
To investigate the benefit of implementing energy storage, we formulate and study the following Peak-demand Reduction Maximization~(\proname) problem, %\com{YF}{consistent with the title}
\begin{equation}
\begin{split}\label{ProPRAD}
  \textbf{\proname:}~~~\max_{\bm \delta \in \mathbb{R}^T} &~~~~\max_{t\in [T]}d_t - \max_{t\in[T]}\left(d_t-\delta_t\right)\\
  \text{subject to} &~~~~\sum_{t=1}^{T}\delta_t\leq c;~~\text{(Inventory Constraint)}\\
  &~~~~0\leq \delta_t \leq \min\{\bar{\delta},d_t\}\text{, for all }t\in [T].
\end{split}
\end{equation}
The objective represents the peak-demand reduction introduced by the energy storage. If there is no storage, the peak usage under the demand profile~$\bm d$ is~$\max_{t\in [T]}d_t$. After applying the discharging vector~$\bm \delta$, we reduce the peak usage to~$\max_{t\in[T]}\left(d_t-\delta_t\right)$. The discharging vector should satisfy the capacity and discharging rate constraints. Moreover, each discharging quantity should not exceed the demand of the corresponding slot. Our goal is to maximize the peak-demand reduction by using energy storage, which directly leads to cost reductions on the peak-demand charge. Most existing studies consider the peak minimization objective, namely,~$\min_{\bm \delta}\max_{t\in[T]} \left(d_t-\delta_t\right)$. In contrast, we adopt the peak-demand reduction maximization objective in \proname, because it makes us focus on the benefit brought by the energy storage. We observe that the two objectives are consistent in the offline setting because their summation remains a known constant~$\max_{t\in [T]}d_t$ under any feasible solution. Notwithstanding, we should differentiate them in the online setting where we lack the information of~$\max_{t\in [T]}d_t$, let alone when we evaluate online algorithms by competitive ratios. Moreover, in the online setting, we consider \rev{the} performance guarantee over an uncertainty set of the demand profiles. $\max_{t\in [T]}d_t$ is non-constant among different demand profiles. Considering a relative peak-reduction value to $\max_{t\in [T]}d_t$ instead of the \rev{absolute} peak-demand value $\max_{t\in[T]}\left(d_t-\delta_t\right)$ provides a more fair performance comparison among different  demand profiles. To the best of our knowledge, our work is the first to study the peak-demand reduction maximization by online competitive analysis.

\subsection{Optimal Offline Solution}\label{sec:offline}
Clearly, when the demand profile~$\bm d$ is known, we can easily solve the \proname~problem by linear programming. Moreover, the following proposition states that the offline optimal solution to \proname~presents a particular threshold-based structure.%, proven in Appendix~\ref{app_offopt},
\begin{proposition}\label{offopt}
  Given a demand profile~$\bm d\in \mathbb{R}^T$, there exists~$v\in \mathbb{R}$ with~$\sum_{t=1}^{T}[d_t-v]^+=c$ and an optimal solution to \proname~is given by
  $$\delta_t^*=\left[d_t-\left[\left[\max_{t\in[T]}d_t-\bar{\delta}-v\right]^++v\right]^+\right]^+\text{, for all }t\in [T].$$
\end{proposition}
Note that $[x]^+\triangleq \max\{x,0\}$. %If we ignore the rate constraints, then the above formula reduces to~$\delta_t^*=[d_t-[v]^+]^+$, for all~$t\in[T]$. Thus, we can regard~$[v]^+$ as a threshold value, determining the discharging quantity in comparison with the net demand at each time slot. If we respect the rate constraints, then the threshold value will be the maximum of~$[v]^+$ and~$[\max_{t\in[T]}d_t-\bar{\delta}]^+$.
%
%Another
A useful observation is that the optimal solution to \proname, is also optimal for the peak-demand minimization problem with the same constraints. For notational convenience, we respectively use~$v(\bm d)$ and~$\sigma(\bm d)$ to denote the optimal peak usage and the optimal peak-demand reduction after discharging stored energy, namely,~$$v(\bm d)=\max_{t\in[T]}(d_t-\delta_t^*)\text{ and }\sigma(\bm d)=\max_{t\in[T]}d_t-\max_{t\in[T]}(d_t-\delta_t^*),$$ where~$\bm \delta^*$ is the optimal solution to \proname. While solving the offline \proname~problem is easy, it is challenging to determine the discharging quantities in real-time without knowing future demands, i.e., in online manner. \cite{bar2008peak} proposed an optimal competitive online algorithm for the peak minimization, assuming that the original peak usage~$\max_{t\in[T]} d_t$ is known a priori. However, such an assumption is far from practical. In this paper, we consider a more general model where the prior information on the peak usage comprises the lower and upper demand bounds only. Also, we propose, in the next section, an optimal competitive algorithm for the online \proname~problem. Overall, our work complements the literature by considering the peak-demand reduction maximization and not assuming to know the exact value of the original peak usage.

\section{Online Algorithm}\label{Sec_OnlineAlg}
In this section, we shall propose the first competitive online algorithm for \proname. This algorithm is parameterized by the best possible CR among all online algorithms and attains the optimal worst-case relative performance guarantee regarding the offline-to-online ratio of the peak reduction. The best possible CR varies as we change the time-slot number~($T$), the demand bounds~($\underline{d}$ and~$\overline{d}$), the storage capacity~($c$), and the maximum discharging limit~($\bar{\delta}$). In general, computing the best CR involves hard min-max optimization and we have to resort to dynamic programming which is time-consuming and computationally expensive. Fortunately, as a unique technical contribution, we show that we can obtain the best CR for the online {\proname} by solving a linear number of linear-fractional programs in parallel. Note that the best possible CR essentially captures the price of uncertainty before we know any inputs, while the price of uncertainty may change as we observe more data and implement actions in past time slots. In practice, we intend to adaptively exploit the real-time information and pursue the updated prices of uncertainty. The particular structure of the proposed algorithm enables us to achieve this purpose and combine efficiency with robustness. We shall elaborate on this extension in the next section.

%%%%%%%%%%%%%%%%%%%%%%%%%%%%%%%%%%%%%%%%%%%%%%%%%%%%%%%%%%%%%%%%%%%%%%%%%%%%%%%%%%%%%%%%%%%%%
\subsection{Online Setting}
\begin{figure*}[t]
  \centering
  \centerline{\includegraphics[width=1.4\columnwidth]{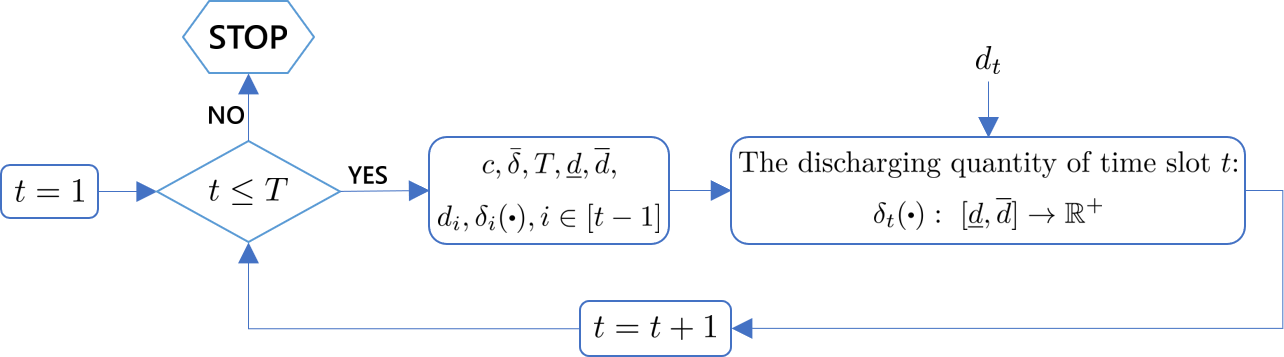}}
  \caption{A flowchart of an online algorithm for \proname.}\label{fig: FlowGraph}
\end{figure*}

In the online setting, we do not know the exact value of~$d_t$ until time slot~$t$. The empirical information only tells us that~$d_t\in[\underline{d},\overline{d}]$, for all~$t\in[T]$. We denote the set of all possible demand profiles by~$\mathcal{D}=\{\bm d\in \mathbb{R}^T \mid \underline{d}\leq d_t \leq \overline{d}, \forall t\in[T]\}$. Other prior information includes the storage capacity~$(c)$, the maximum discharging limit~$\bar{\delta}$, and the time-slot number~$(T)$. We present, in Fig.~\ref{fig: FlowGraph}, an illustrative flowchart of an online algorithm for \proname.

As mentioned in the introduction, concerning robustness and fairness, CR is a proper measure in online optimization with given resources~\citep{borodin2005online}. It quantifies the relative performance between an online algorithm and the offline optimal benchmark. For a maximization problem like \proname, the CR of a deterministic online algorithm~$\mathfrak{A}$ is defined as the largest ratio between the objective value under an optimal offline solution and that attained by the algorithm over all possible input sequences~(e.g., the demand profiles in \proname), namely,
\begin{equation*}
  CR_{\mathfrak{A}}=\max_{\bm d\in \mathcal{D}}\frac{\sigma(\bm d)}{\sigma_{\mathfrak{A}}(\bm d)},
\end{equation*}
where~$\sigma_{\mathfrak{A}}(\bm d)$ refers to the objective values of {\proname} under the online algorithm and the input sequence~$\bm d$. For a randomized algorithm~$\mathfrak{A}$, we can extend the concept of CR by replacing~$v_{\mathfrak{A}}(\bm d)$ with~$\mathbf{E}[v_{\mathfrak{A}}(\bm d)]$, where the expectation is due to the random strategies of the algorithm~$\mathfrak{A}$. We always have~$CR_{\mathfrak{A}}\geq 1$ and a smaller CR indicates that the algorithm can perform more closely to the offline optimal case. We expect to find the online algorithm with the smallest CR, which is the best possible CR among all online algorithms. The best CR not only verifies the optimal performance of the online algorithm, but also quantifies the essential cost of not knowing the future. In other words, the best CR captures the price of uncertainty in the considered online problem.

%Intuitively, with a larger design space, a randomized algorithm can attain a better CR than its deterministic counterpart. However, t
In the following, we show that it suffices to focus on deterministic online algorithms for the best CR regarding the online {\proname}.%, proven in Appendix~\ref{app_rand},
\begin{proposition}\label{prop_rand}
  For any randomized algorithm~$\mathfrak{A}$, there exists a deterministic online algorithm~$\mathfrak{B}$ such that~$\mathbf{E}[\sigma_{\mathfrak{A}}(\bm d)]\leq \sigma_{\mathfrak{B}}(\bm d)$ over all possible input sequence~$\bm d\in \mathcal{D}$.
\end{proposition}

%%%%%%%%%%%%%%%%%%%%%%%%%%%%%%%%%%%%%%%%%%%%%%%%%%%%%%%%%%%%%%%%%%%%%%%%%%%%%%%%%%%%%%%%%%%%%
\subsection{Overview of the \textsf{CR-Pursuit} framework}\label{Sec_CRP}
The usual practice is to propose an online algorithm and then compute its CR for evaluating the performance. In contrast, we herein ask whether there exists an online algorithm with given CR. This question relates to a useful framework for designing competitive online algorithms, called~\textsf{CR-Pursuit}. As the name indicates, the~\textsf{CR-Pursuit} algorithmic framework requires us to sequentially make online decisions by ``pursuing'' a prescribed CR. For a maximization problem and a given CR, each~\textsf{CR-Pursuit} algorithm will choose actions to maintain an offline-to-online objective ratio to be no more than the CR, in each decision-making round. Despite the conciseness of the idea, there is no general recipe for pursuing the given CR. The means of designing algorithms under~\textsf{CR-Pursuit} is problem-specific, as exemplified in~\cite{lin2019competitive} and~\cite{yi2019balancing}.

We herein summarize several useful observations regarding the~\textsf{CR-Pursuit} framework. First, to maintain the offline-to-online ratio in each decision-making round, we usually need to solve an offline problem according to the observed inputs and implemented actions so far. If the offline version of a considered problem is easier, then it is less challenging to design algorithms under~\textsf{CR-Pursuit}. Second, the~\textsf{CR-Pursuit} framework generally generates a family of algorithms, each of which is characterized by a specific CR. Clearly, the best algorithm among these is the one with the smallest CR. We wonder whether the best algorithm under~\textsf{CR-Pursuit} is optimal among all online algorithms. If so, then the~\textsf{CR-Pursuit} framework greatly reduces the search space of optimal online algorithms. Overall, there are three main challenges in designing~\textsf{CR-Pursuit} algorithms: finding a proper way to sequentially maintain a given CR, identifying the best CR which can be pursued, and checking whether the best CR under~\textsf{CR-Pursuit} is the optimal one among all online algorithms. In the following, we shall tackle these challenges with the online {\proname}. First of all, we shall devise a collection of online algorithms under the~\textsf{CR-Pursuit} framework. Then, we show that the best algorithm among these is also optimal among all online algorithms for {\proname}. Finally, we attain the best online algorithm for {\proname} by finding the best possible CR.
%%%%%%%%%%%%%%%%%%%%%%%%%%%%%%%%%%%%%%%%%%%%%%%%%%%%%%%%%%%%%%%%%%%%%%%%%%%%%%%%%%%%%%%%%%%%%
\subsection{Optimal Online Algorithm for {\proname}}
Without knowing future inputs, it seems impossible to maintain the offline-to-online objective ratio to be no more than a given ratio. If we do not know the time-slot number~$(T)$, we can simply assume that there are no future inputs, as has been done in the existing results~\citep{lin2019competitive,yi2019balancing}. However, in the online {\proname}, we are challenged by the facts that~$T$ is known and the future inputs will affect the online and offline optimal peak-demand reductions. As a result, we shall introduce a reference input and adaptively update it at each time slot~$t\in[T]$. Specifically, we set the reference input sequence at time slot~$t$ as
$$\bm d^t=[d_1~d_2~\cdots~d_t~\underline{d}~\cdots~\underline{d}]'.$$
Clearly, the reference input sequence is a combination of the observed demands until time slot~$t$ and the lowest possible future demands, indicating the most optimistic forecast. Given a ratio~$\pi$, we shall apply~\textsf{CR-Pursuit} framework and maintain the offline-to-online ratio under the reference input sequence~$\bm d^t$ to be more than~$\pi$ at each time slot~$t\in[T]$. To this end, we design a family of algorithms, each of which is characterized by a prescribed ratio~$\pi$ and called~\algname($\pi$). The details of~\algname($\pi$) are given in Algorithm~\ref{algpCS}. For notational convenience, we use~$\bm \delta(\pi,\bm d)$ to denote the output sequence of~\algname($\pi$) under input sequence~$\bm d$. %By the formula in Algorithm~\ref{algpCS}, we see that
%$$\max_{k\in[t]} d_k-(d_t-\delta_t(\pi,\sigma))\geq \sigma(\bm d^t)/\pi\text{, for all }t\in[T].$$

\begin{algorithm}[t]
\caption{\algname($\pi$)}\label{algpCS}
\LinesNumbered
  %\begin{algorithmic}[1]
  \For{$t=1,2,\ldots,T$} {
  {The discharging quantity at time slot~$t$ is given by~$\delta_t(\pi,\bm d) = [d_t -\max_{k\in[t]}d_k+\sigma(\bm d^t)/\pi]^+$;}}
  %\end{algorithmic}
\end{algorithm}

Alert readers may put forward two issues. First,~\algname($\pi$) may generate infeasible solutions to {\proname}. Second, we wonder whether the formula in Algorithm~\ref{algpCS} can maintain the offline-to-online objective ratio under~$\bm d^t$ to be no more than~$\pi$, for all~$t\in[T]$. For the first issue, we give the following definition in terms of feasibility of~\algname~algorithms.
\begin{definition}
  \algname($\pi$) is feasible if, for any~$\bm d\in \mathcal{D}$, the solution~$\bm \delta(\pi,\bm d)$ is feasible for {\proname}.
\end{definition}
We address the second issue by the following lemma.%, proven in Appendix~\ref{app_maintain}.
\begin{lemma}\label{lemmaintain}
Given~\algname($\pi$), it holds for any~$\bm d\in\mathcal{D}$ that
$$\max_{k\in[t]} d_k-\max_{k\in[t]}(d_k-\delta_k(\pi,\bm d))\geq \sigma(\bm d^t)/\pi\text{, for all } t\in[T].$$
\end{lemma}

Therefore, we conclude that~\algname($\pi$) can maintain the given CR~$\pi$ if and only if it is feasible. Our goal is to find the smallest~$\pi$ such that~\algname($\pi$) is feasible. To this end, we shall first characterize the set of all ratios such that the corresponding \algname~algorithms are feasible. For this purpose, we define an inventory function, which maps a given ratio~$\pi$ to the maximum accumulated discharging quantity over all possible demand profiles under~\algname($\pi$):
$$\Phi(\pi)=\max_{\bm d\in \mathcal{D}}\sum_{t=1}^{T}\delta_t(\pi,\bm d).$$
We show below that the feasibility of~\algname($\pi$) is mainly subject to the inventory constraint.
\begin{proposition}\label{propfeasibility}
  \emph{\algname($\pi$)} is feasible if and only if~$\Phi(\pi)\leq c$.
\end{proposition}
%We prove Proposition~\ref{propfeasibility} in Appendix~\ref{app_feasibility}.
The following lemma unravels the monotonicity of the inventory function~$\Phi(\pi)$.%, proven in Appendix~\ref{app_deinc},
\begin{lemma}\label{lemdeinc}
  $\Phi(\pi)$ is non-increasing and strictly decreases in $\pi$ when $\Phi(\pi)>0$.
\end{lemma}

From Lemma~\ref{lemdeinc}, we see that there exists a ratio~$\bar{\pi}>1$ such that~\algname($\pi$) is feasible only if~$\pi\geq \bar{\pi}$. We shall show that~\algname($\bar{\pi}$) is the best~\algname~algorithm. More importantly, the best~\algname~algorithm also attains the optimal CR among all online algorithms for {\proname}, as stated in the following theorem.
\begin{theorem}\label{thmoptratio}
  Given~$c$,~$\bar{\delta}$,~$T$, and~$\mathcal{D}$,  the unique solution~$\pi^*$ to the equation~$\Phi(\pi)=c$ is the best possible competitive ratio among all online algorithms for {\proname}.
\end{theorem}

The above theorem indicates that the solution to~$\Phi(\pi)=c$ characterizes the price of uncertainty regarding the online {\proname}. To find the best online algorithm for {\proname}, it suffices to search for the best feasible~\algname~algorithm. In the next subsection, we shall show an efficient way to find the best possible CR~$\pi^*$.%, without knowing the explicit form of the complicated function~$\Phi(\pi)$. The best CR~$\pi^*$ varies as we change the time-slot number ($T$), the demand bounds~($\underline{d}$ and~$\overline{d}$), the storage capacity~($c$), and the maximum discharging limit~($\bar{\delta}$). In this case, computing the best CR usually involves hard min-max optimization and we have to resort to dynamic programming which is time-consuming and computationally expensive.

%%%%%%%%%%%%%%%%%%%%%%%%%%%%%%%%%%%%%%%%%%%%%%%%%%%%%%%%%%%%%%%%%%%%%%%%%%%%%%%%%%%%%%%%%%%%%
\subsection{Finding the Optimal Competitive Ratio~$\pi^*$}
%We reserves it as an inspiring future direction to derive the analytic expression of~$\pi^*$ in terms of the four-tuple of parameters~$(c,T,\underline{d},\overline{d})$. Actually, it does not seem easy to obtain the numerical value of~$\pi^*$ with given~$(c,T,\underline{d},\overline{d})$, owing to the fact
%We see that the optimal CR is a function over~$(c,T,\mathcal{D})$, written as~$\pi^*(c,T,\mathcal{D})$.
Deriving the analytic expression of~$\pi^*$ over all the parameters $(c,\bar{\delta},T,\underline{d},\overline{d})$ is attractive but challenging. Instead, we shall explore efficient numerical methods for the best CR~$\pi^*$. As a unique technical contribution, we show that computing~$\pi^*$ is not much harder than linear programming. Specifically, we first show that the best CR~$\pi^*$ is the maximum of an exponential number of linear-fractional programs. Then, we exploit the problem structure and show that only a linear number of such programs are necessary in terms of the time-slot number~$T$. Note that we can transform each involved linear-fractional program into an equivalent linear program by the techniques in~\cite[Section 4.3.2]{boyd2004convex}.

Recall the key formula of~\algname($\pi$) and the feasibility condition~$\Phi(\pi)\leq c$ of~\algname($\pi$). We conclude that~\algname($\pi$) is feasible if and only if for any nonempty subset~$\mathcal{I}$ of the index set~$[T]$ and any demand profile~$\bm d\in \mathcal{D}$, it holds that~$\sum_{i\in \mathcal{I}}[d_i -\max_{k\in[i]}d_k+\sigma(\bm d^i)/\pi]^+\leq c$. It follows that
\begin{equation}\label{lfinequality}
 \frac{\sum_{i\in \mathcal{I}}\sigma(\bm d^i)}{c+\sum_{i\in \mathcal{I}}\left(\max_{k\in[i]}d_k - d_i\right)}\leq \pi.
\end{equation}
A worst-case input sequence for~\algname($\pi^*$) refers to a demand profile~$\bm d\in \mathcal{D}$, under which the~\algname($\pi^*$) algorithm will use up the storage, namely~$\sum_{t=1}^{T}\delta(\pi^*,\bm d)=c$. Since~$\Phi(\pi^*)=c$, there exists a worst-case input sequence for~\algname($\pi^*$). Let~$\bm d^*$ be a worst-case input sequence and~$\mathcal{I}^*$ be the set~$\{i\in[T]\mid \delta_t(\pi^*,\bm d^*)>0\}$. Then, we observe that the equality in Formula~(\ref{lfinequality}) holds under the index set~$\mathcal{I}^*$ and the input sequence~$\bm d^*$. Therefore, we attain the following proposition.
\begin{proposition}\label{proppaiexp}
  Given~$c$,~$\bar{\delta}$,~$T$, and~$\mathcal{D}$, the best possible CR for the online {\proname} is given by~$$
\pi^*=\max_{\mathcal{I}\subseteq [T],\bm d\in \mathcal{D}}\frac{\sum_{i\in \mathcal{I}}\sigma(\bm d^i)}{c+\sum_{i\in \mathcal{I}}\left(\max_{k\in[i]}d_k - d_i\right)}.
$$
\end{proposition}
Proposition~\ref{proppaiexp} follows directly from Lemma~\ref{lemdeinc} and Proposition~\ref{propfeasibility}. Equipped with Proposition~\ref{proppaiexp}, we shall transform the computation of the best CR~$\pi^*$ into solving a sequence of linear-fractional programs. Before proceeding, we observe that~$\sigma(\bm d)=\max_{t\in[T]}d_t-\bm v(\bm d)$ and the largest element in~$\bm d^t$ appears in the first~$t$ elements. Moreover, we introduce a useful lemma below.
\begin{lemma}\label{lemaddconst}
  Given~$x>y>0$ and~$z>0$, it follows that~$\frac{x+z}{y+z}\leq \frac{x}{y}$.
\end{lemma}

Next, we shall define a family of linear-fractional programs. Each program is parameterized by a nonempty set~$\mathcal{I}\subseteq [T]$ and its optimal objective value gives a lower bound of the best CR~$\pi^*$:%and qualitatively characterize the relationship between~$\pi^*$ and
\begin{equation*}
\begin{split}
  \textsf{CR-Comp}(\mathcal{I}):~&\max_{\bm d\in \mathcal{D},u_i,\delta_{ij}}~~\frac{\sum_{i\in\mathcal{I}}(m_i-u_i)}{c+\sum_{i\in \mathcal{I}}(m_i-d_i)}\\
      \text{subject to}~&\sum_{j=1}^{T}\delta_{ij}\leq c\text{, for all }i\in[T];\\
      & 0\leq \delta_{ij}\leq \bar{\delta}\text{, for all }i,j\in [T];\\
      & d_j-\delta_{ij}\leq u_i\text{, for all }1\leq j\leq  i\leq T;\\
      & \underline{d}-\delta_{ij}\leq u_i\text{, for all }1\leq i< j\leq T;\\
      & d_k\leq m_i\text{, for all }k\in[i] \text{ and }i\in \mathcal{I}.
    \end{split}
  \end{equation*}
%  \com{yf}{specify that~$m_i$ is an auxiliary variable and its use }
Now, let us interpret the variables, constraints and objective of~\textsf{CR-Comp}($\mathcal{I}$). $m_i,i\in[T]$ are auxiliary variables corresponding to $\max_{k\in[i]}d_k$. More specifically, by Lemma~\ref{lemaddconst} and the constraints on the last line,~\textsf{CR-Comp}($\mathcal{I}$) will attain its optimum when~$m_i=\max_{k\in[i]}d_k$, for all~$i\in\mathcal{I}$. The remaining constraints of~\textsf{CR-Comp}($\mathcal{I}$) are associated with the offline {\proname} problem solved at each time slot under a~\algname~algorithm. Precisely, for each~$i\in[T]$, the variable~$v_i$ and the variables~$\delta_{ij},j\in[T]$ are respectively related to the optimal objective value and the optimal solution to {\proname} under the demand profile~$\bm d^i$. \textsf{CR-Comp}($\mathcal{I}$) will attain its optimum when~$v_i=v(\bm d^i)$, for all~$i\in\mathcal{I}$. Moreover, the objective function of~\textsf{CR-Comp}($\mathcal{I}$) corresponds to the left part of Formula~(\ref{lfinequality}), noting that $\sigma(\bm d^i)=\max_{k\in[i]}d_k-v(\bm d^i)$. With a slight abuse of notation, we also use~$\textsf{CR-Comp}(\mathcal{I})$ to denote the optimal objective value of the linear-fractional program. As a whole, by Lemma~\ref{lemaddconst}, Proposition~\ref{proppaiexp}, and the above analysis, we attain the following proposition.
\begin{proposition}\label{proptolf}
  For each nonempty set~$\mathcal{I}\subseteq [T]$,~$\textsf{CR-Comp}(\mathcal{I})=\max_{\bm d\in \mathcal{D}}\frac{\sum_{i\in \mathcal{I}}\sigma(\bm d^i)}{c+\sum_{i\in \mathcal{I}}\left(\max_{k\in[i]}d_k - d_i\right)}.$
\end{proposition}

Thus, by Proposition~\ref{proppaiexp} and Proposition~\ref{proptolf}, we conclude that the best possible CR~$\pi^*$ equals~$\max_{\mathcal{I}\subset [T]}\textsf{CR-Comp}(\mathcal{I})$. That is to say, we can compute~$\pi^*$ by solving a collection of linear-fractional programs~\textsf{CR-Comp}($\mathcal{I}$). Recall that we can convert~\textsf{CR-Comp}($\mathcal{I}$) into an equivalent linear program. Thus, computing~$\pi^*$ is not much harder than solving a collection of linear programs. Moreover, we can compute these programs in parallel, since none of the $\textsf{CR-Comp}(\mathcal{I})$ programs relies on the solution to another.  Nevertheless, a direct application of the above results requires to solve an exponential number of linear programs, which is undesirable. Thus, we are motivated to exploit the structure of~\algname($\pi^*$) and exclude as many redundant programs as possible. To this end, we derive the following lemma. It identifies a particular worst-case input sequence for~\algname($\pi^*$), which continuously discharge stored energy until using up the capacity.

\begin{figure}[t]
  \centering
  \begin{subfigure}[t]{0.49\columnwidth}
    \includegraphics[width=\columnwidth]{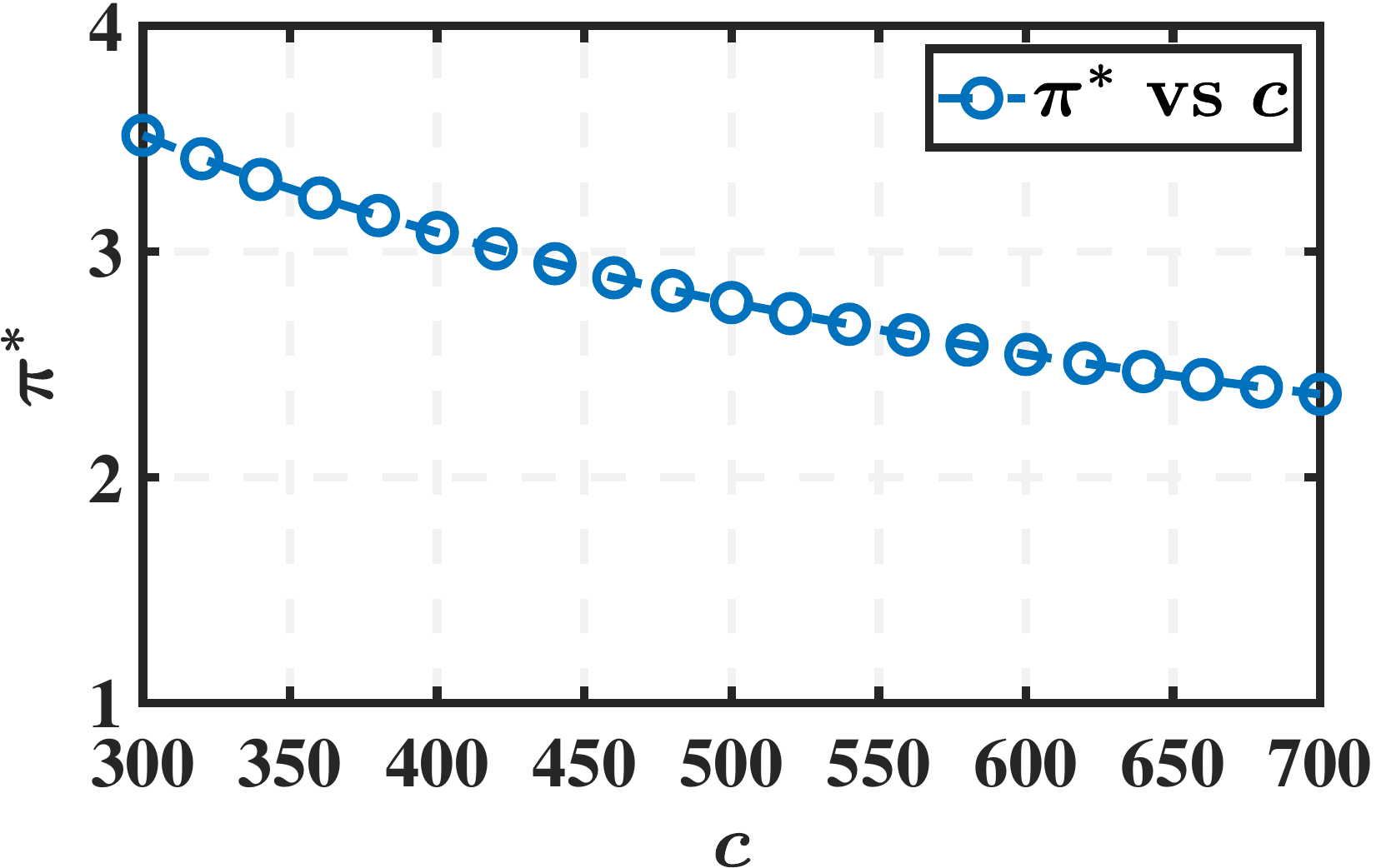}
  \end{subfigure}
  \hfill
  \begin{subfigure}[t]{0.49\columnwidth}
    \includegraphics[width=\columnwidth]{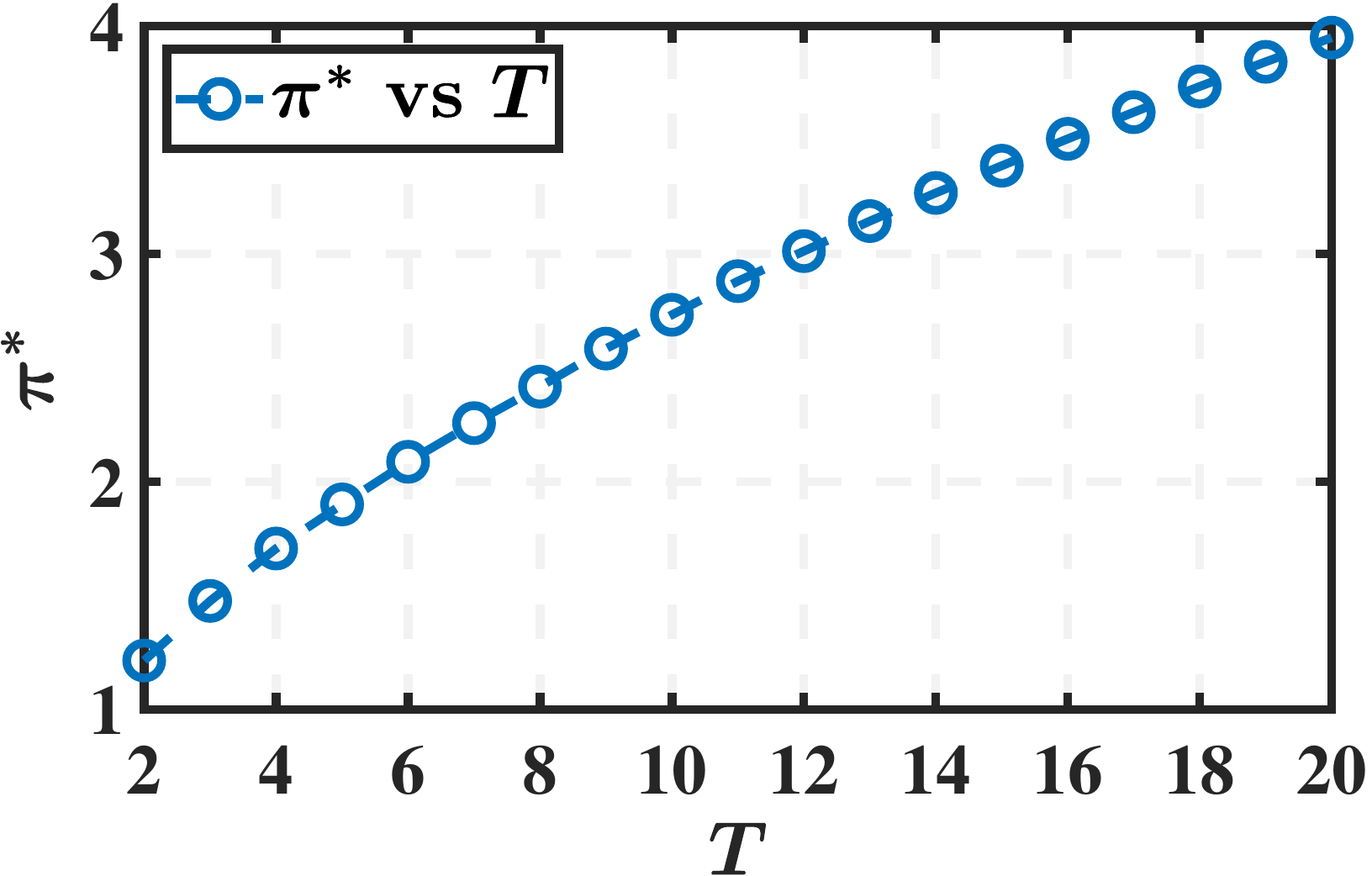}
  \end{subfigure}
  \vskip\baselineskip
  \begin{subfigure}[t]{0.49\columnwidth}
   \includegraphics[width=\columnwidth]{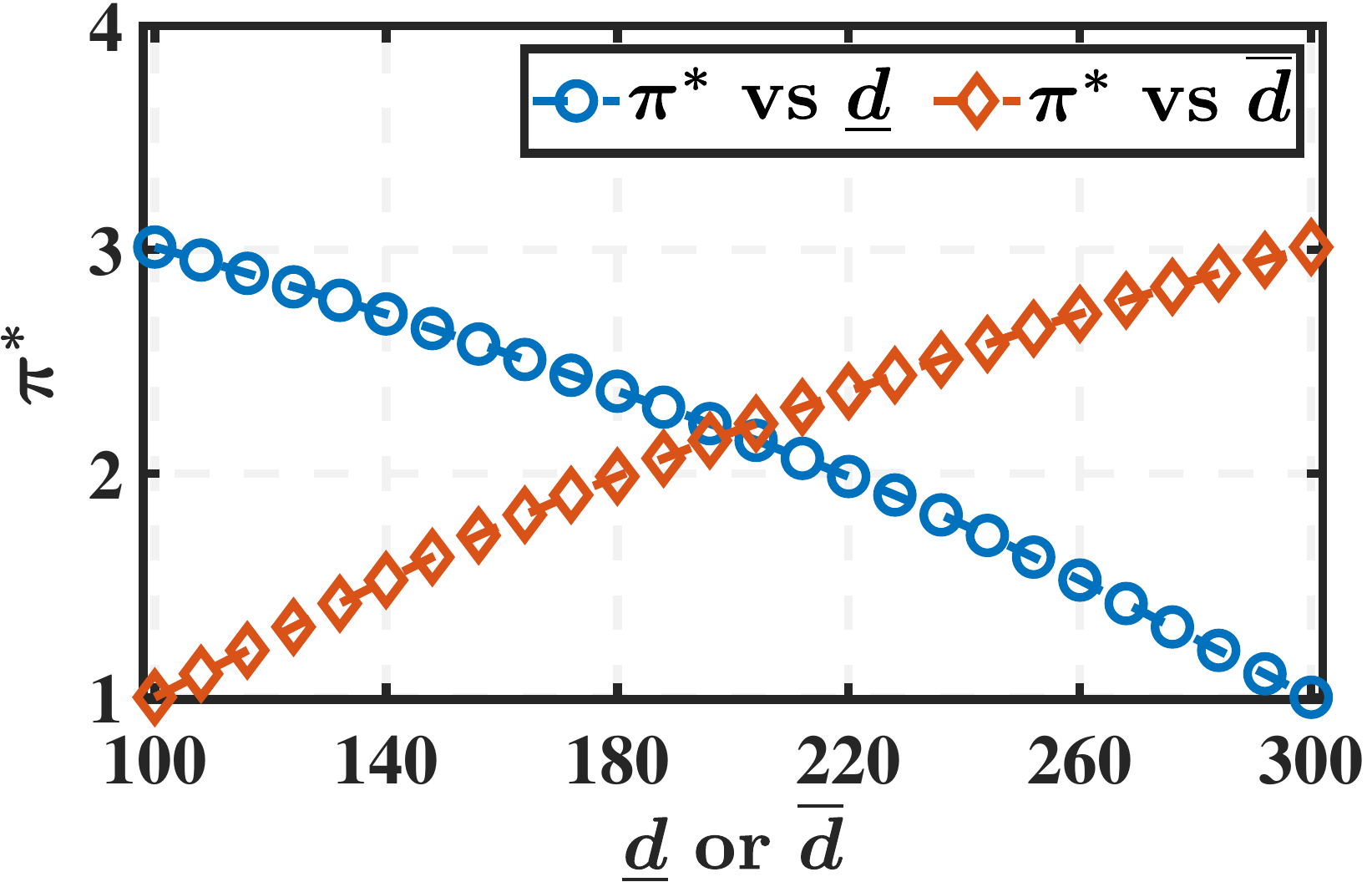}
  \end{subfigure}
  \hfill
   \begin{subfigure}[t]{0.49\columnwidth}
    \includegraphics[width=\columnwidth]{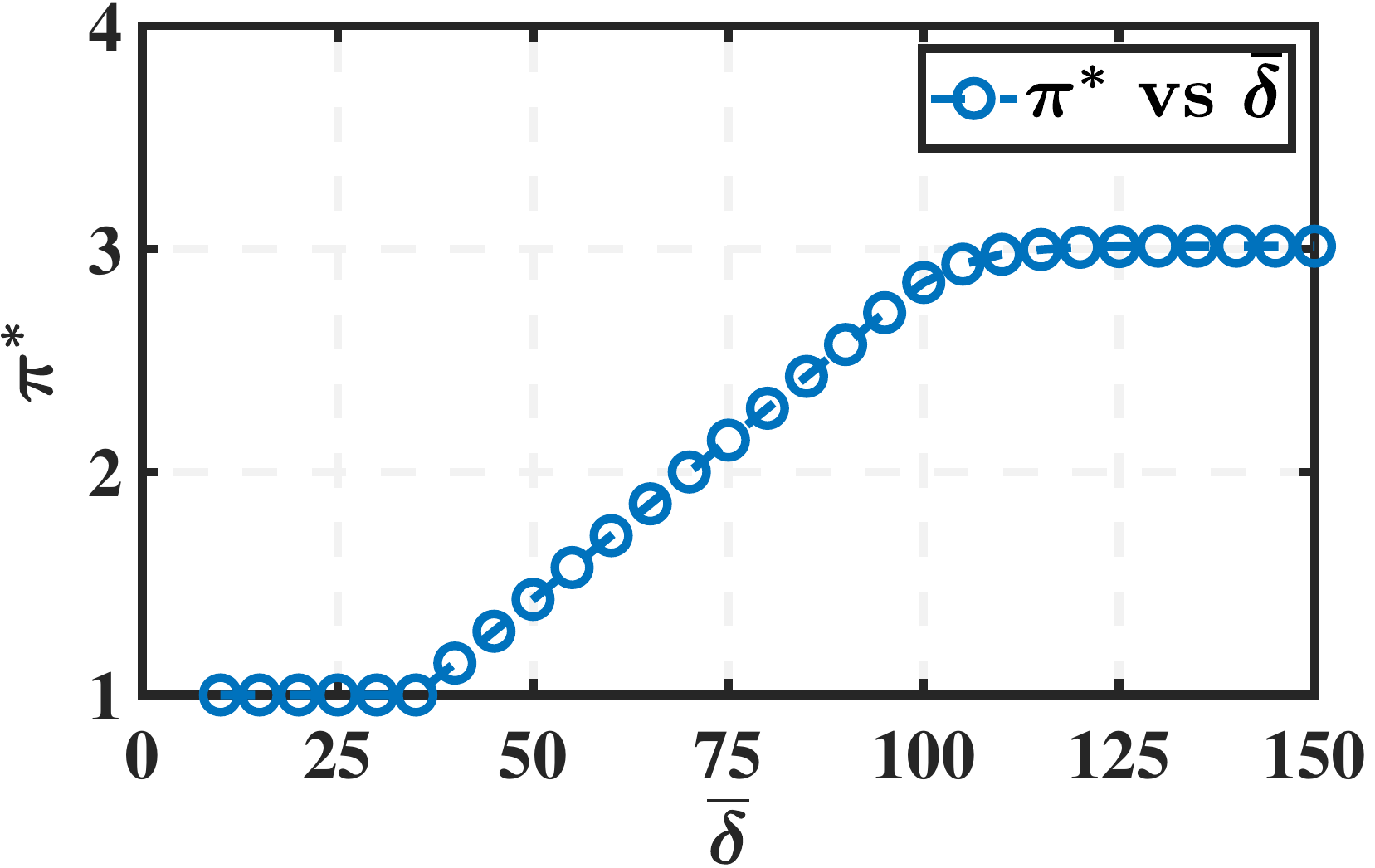}
  \end{subfigure}
%  \begin{subfigure}[t]{0.23\textwidth}
%    \leftline{\includegraphics[width=1.2\columnwidth]{figures/pi_lb_ub_delta-eps-converted-to.pdf}}
%  \end{subfigure}
  \caption{An illustration on how~$\pi^*$ varies as we vary the parameters~$(c,\bar{\delta},T,\underline{d},\overline{d})$. The default values are set as~$c=420$,~$T=12$,~$\underline{d}=100$,~$\overline{d}=300$,  and~$\bar{\delta}=300$.}
  \label{fig: Sensitivity}
\end{figure}

%Proposition~\ref{proppaicomb} means that there exists a set~$\mathcal{I}\subset [T]$ such that the associated~\textsf{CR-Compute}($\mathcal{I}$) just equals the optimal CR~$\pi^*$.
%Theorem~\ref{proppaicomb} suggests a numerical way to obtain the optimal CR~$\pi^*$.

\begin{lemma}\label{lemparwc}
There exists a worst-case input sequence~$\bm d\in \mathcal{D}$ for~\algname($\pi^*$) such that~$\delta_t(\pi^*,\bm d)=0$ if~$t$ is greater than a certain index and~$\delta_t(\pi^*,\bm d)>0$ otherwise.
\end{lemma}

%We prove the above lemma in Appendix~\ref{app_parwc}.
\rev{By Proposition~\ref{proppaiexp}, Proposition~\ref{proptolf}, and Lemma~\ref{lemparwc}, we see that there exists~$k\in[T]$ such that~$\pi^*=\textsf{CR-Comp}([k])$. Thus ,we have the following theorem stating that only a linear number~($T$) of~\textsf{CR-Comp}($\mathcal{I}$) programs are necessary for the best possible CR~$\pi^*$, instead of the exponential number~$2^T$.} % the discharging will never happen in the remaining time slots once~\textsf{pCR-PAD}($\pi^*$) decides not to discharge in a certain time slot
\begin{theorem}\label{thmoptratiocal}
    The best possible competitive ratio~$\pi^*$ for the online {\proname} is given by %the optimal objective value of the following optimization problem:
  \begin{equation*}
 ~~~\pi^*~=~\max_{\mathcal{I}\in\{[t]\mid t\in [T]\}}~ \textsf{CR-Comp}(\mathcal{I}).
\end{equation*}
\end{theorem}
%\begin{proof}
%  By Lemma~\ref{lemparwc}, there exist an input sequence~$\bm d$ and a index~$i\in[T]$, such that
%  \begin{equation*}
%    \sum_{t=1}^{i}\delta_t(\pi^*,\bm d)=c \text{ and } \delta_t(\pi^*,\bm d)>0
%    \text{, for all } t\in [i].
%  \end{equation*}
%  It follows that
%  \begin{equation*}
%    \frac{\sum_{t=1}^{i}d_t-c}{\sum_{t=1}^{i}v(\bm d^t)}=\pi^*.
%  \end{equation*}
%  Applying Proposition~\ref{proppaicomb}, we show that
%  \begin{equation*}
% ~~~\pi^*~=~\max_{\mathcal{I}\in\{[t]\mid t\in[T]}~ \textsf{CR-Compute}(\mathcal{I}).
%\end{equation*}
%Finally, we complete the proof by the fact that~$\pi^*\geq 1$ while~\textsf{CR-Compute}$([i])\leq 0$ for all~$i\leq \tau$.
%\end{proof}

In Fig.~\ref{fig: Sensitivity}, we illustrate how the best possible CR~$\pi^*$ varies as each of the parameters~$(c,\bar{\delta},T,\underline{d},\overline{d})$ changes.
We observe that the optimal CR $\pi^*$ increases as $T$ increases and $c$ decreases. $\pi^*$ becomes smaller when $\underline{d}$ and $\overline{d}$ getting closer as it reduces the space of the demand profiles. $\pi^*$ is non-decreasing in $\bar{\delta}$. When $\bar{\delta}$ is small, e.g.,~$\bar{\delta}\leq c/T$, the best reduction for both online and offline is $\bar{\delta}$, and the optimal CR is one.
%Interpret the information from figures????.

%%%%%%%%%%%%%%%%%%%%%%%%%%%%%%%%%%%%%%%%%%%%

%%%%%%%%%%%%%%%%%%%%%%%%%%%%%%%%%%%%%%%%%%%%%%%%%%%%%%%%%%%%%%%%%%%%%%%%%%%%%%%%%%%%%%%%%%%%%
\section{Adaptive~\algname~ Algorithm}\label{Sec_Adaptive}
%%%%%%%%%%%%%%%%%%%%%%%%%%%%%%%%%%%%%%%%%%%%%%%%%%%%%%%%%%%%%%%%%%%%%%%%%%%%%%%%%%%%%%%%%%%%%
%\subsection{}\label{Sec_Adaptive}
While~\algname($\pi^*$) attains the optimal CR among all online algorithms for {\proname}, it merely focuses on the worst-case performance, which may restrict its performance in practice. We herein extend the~\algname($\pi^*$) algorithm by adaptively exploiting the revealed information of previous slots. Here is the intuitive idea: when we realize from the observed inputs that the net demand profile is by no means a worst-case input sequence, we should be more opportunistic and attempt to maintain smaller ratios in the following time slots. In this way, we can improve the average-case performance and still attain the optimal worst-case performance. The underlying reason lies in that the price of future uncertainty changes as we observe more inputs and capturing such variations is critical to the improvement of online decisions. Overall, the results in this section suggest the potential of merging efficiency into robustness. %Moreover, the adaptive extension unravels the opportunities to decrease the volume charge in a bill further but not interfere with the peak-demand charge reduction in terms of competitive analysis.

With respect to the real-time information, we first extend the concept of CR to a time-variant adaptive CR for every online algorithm for {\proname} at each time slot $t$. Specifically, we make the following definition.

\begin{definition}\label{defn:adapCR}
Given revealed inputs~$d_k,k\in[t]$, the adaptive CR at time slot~$t$ of an online algorithm~$\mathfrak{A}$ is defined as
\[\pi^\mathfrak{A}_t = \max_{\bm x\in \mathcal{D}~\&~x_k=d_k\text{, for all }k\in[t]}\frac{\sigma(\bm x)}{\sigma_{\mathfrak{A}}(\bm x)}.\]
\end{definition}
Let~$\mathcal{A}_{t}$ be the set of online algorithms whose first~$(t-1)$ outputs are~$\delta_k,k\in [t-1]$ supposing that the first~$(t-1)$ inputs are~$d_k,k\in [t-1]$. Then, considering the observed inputs~$d_k,k\in [t]$ and implemented actions~$\delta_k,k\in [t-1]$ so far, the best adaptive CR at time slot~$t$ is given by
\begin{equation*}
  \pi^*_t = \min_{\mathfrak{A}\in \mathcal{A}_t}~\pi^\mathfrak{A}_t.
\end{equation*}

\begin{algorithm}[t]
\caption{Adaptive \algname~Algorithm}\label{algadaptivepCR}
\LinesNumbered
  %\begin{algorithmic}[1]
  \For{$t=1,2,\ldots,T$}{
  {Obtain $\pi_t^*$ according to Algorithm~\ref{algadaptivepCRcompute},}

{The discharging quantity at time slot~$t$ is given by
% \[
%           \delta_t = \begin{cases}
%     [d_t - \pi^*_t  \cdot v(\bm d^t)]^+ &\text{if } \pi^*_t \cdot v(\bm d^t)> \max_{k\in [t-1]} (d_k-\delta_k);\\
%     [d_t - \max_{k\in [t-1]} (d_k-\delta_k)]^+ &\text{otherwise}.%
% \end{cases}
% \]
\[
\delta_t(\pi,\bm d) =\left [d_t -\max_{k\in[t]}d_k+\sigma(\bm d^t)/\pi_t^*\right]^+
\]
 }}
 % \end{algorithmic}
\end{algorithm}

Similarly to before, the best adaptive CR at time slot~$t$ characterizes the price of uncertainty at time slot~$t$, for all~$t\in[T]$. Specifically, an online algorithm can at best maintain the online-to-offline ratio of peak-demand reduction to be~$\pi^*_t$, for all future inputs. It is clear that~$\pi^*_t$ is subject to the observed inputs and actions, for all~$t\in[T]$. Based on the introduction of best adaptive CRs, we are ready to introduce the adaptive extension of~\algname.

% algorithm improves~\algname($\pi^*$) as we pursue the best possible adaptive CR at any time slot based on revealed inputs and online decisions so far. That is,
At each time slot~$t$, the adaptive~\algname~ maintains the online-to-offline ratio of the peak-demand reduction to be no more than the best adaptive CR~$\pi^*_t$,  instead of a constant CR $\pi^*$. %since~$\pi^*_t$ essentially captures the price of future uncertainty given the revealed inputs and implemented actions so far.
We present the pseudocodes of the adaptive~\algname~in Algorithm~\ref{algadaptivepCR}. %Note that it is unnecessary to decrease the purchased demand of the current time slot to be less than the peak purchased demand in previous time slots. This fact brings additional difficulty in characterizing the best adaptive CRs, compared to the best CR~$\pi^*$, as elaborated in the sequel.

The remaining issue is on characterizing $\pi_t^*$, the best adaptive CR at each slot. To proceed, we rely on the following observations.

\begin{itemize}
    \item We observe that the best adaptive CR~$\pi^*_1$ is no more than the best CR~$\pi^*$, because we exploit the additional information~$d_1$. Since the adaptive~\algname~algorithm makes decisions by pursuing the best adaptive CR at each time slot, we conclude that the sequence~$\pi^*_t,t\in[T]$ is nonincreasing in~$t$,% i.e., (define $\pi_0^*=\pi^*$),
    \[
    \pi_t^*\leq \pi_{t-1}^*, \forall t\in[T].
    \]
    \item Given the online actions before time slot $t$, $\delta_k,k\in[t-1]$, the online peak-demand reduction under the demand profile $\bm d^t$ is no more than \mbox{$\max_{k\in[t]}d_k-\max_{k\in[t-1]}{\left(d_k-\delta_k\right)}$}, %If the equality holds, we have~$\delta_t=[d_t-\max_{k\in[t-1]}(d_k-\delta_k)]^+$.
    Thus, defining
    \[
    \pi^{lb}_t \triangleq \frac{\sigma(\bm d^t)}{\max_{k\in[t]}d_k-\max_{k\in[t-1]}\left(d_k-\delta_k\right)},
    \]
    we have $\pi^*_t\geq \pi_t^{lb}$.
\end{itemize}

Based on the observations, we shall show how to search for~$\pi^*_t$ by a bisection method. To this end, in the following, given observed inputs~$d_k,k\in[t]$ and made decisions~$\delta_k,k\in [t-1]$, we define a linear program parameterized by a ratio~$\pi\in [\pi_{lb}^t,\pi^*_{t-1}]$ and a set~$\mathcal{I}\in \mathcal{I}_t$, where~$\mathcal{I}_t=\{[k]\backslash [t]\mid k=t,t+1,\ldots,T\}$:

\begin{algorithm}[t]
\caption{A Bisection Method for~$\pi^*_t$ in the Adaptive \algname~ Algorithm}\label{algadaptivepCRcompute}
%\LinesNumbered
  %\begin{algorithmic}[1]
  \KwIn{$c,T,\underline{d},\overline{d}$, observed inputs~$d_k,k\in[t]$, implemented actions~$\delta_k,k\in[t-1]$, and~$\pi^*_{t-1}$;}
  \KwOut{The adaptive CR at time slot~$t$ under Adaptive \algname: $\pi^*_t$;}
  {$\pi_{lb}=\pi_t^{lb}$, $\pi_{ub}=\pi^*_{t-1}$, $q = \max_{\mathcal{I}\in \mathcal{I}_t} \textsf{AdaCR-Threshold}(\pi_{lb},\mathcal{I})$;}

  \If{$q\leq (c-\sum_{k=1}^{t-1}\delta_k)$} {$\pi^*_t=\pi_{lb}$,

  \Return%\textbf{break};
  }

  %\Else{
  \While{$\pi_{ub}-\pi_{lb}\geq \epsilon$}{
  {$\pi=(\pi_{lb}+\pi_{ub})/2$, $q = \max_{\mathcal{I}\in \mathcal{I}_t} \textsf{AdaCR-Threshold}(\pi,\mathcal{I})$;}
   {
  \[ \begin{cases}
\pi_{lb}=\pi & \text{if } q > (c-\sum_{k=1}^{t-1}\delta_k);\\
    \pi_{ub}=\pi &\text{otherwise};
\end{cases}
\]}
  %\IF{$q > (c-\sum_{k=1}^{t-1}\delta_k)$}
%  \STATE{$\pi_{lb}=\pi$;}
%  \ELSE \STATE{$\pi_{ub}=\pi$;}
%  \ENDIF
  }
 {$\pi^*_{t}=\pi_{ub}$;}
  %}
  %\end{algorithmic}
\end{algorithm}

%   \begin{equation*}
%     \begin{split}
%     \textsf{AdaCR-Threshold}(\pi,\mathcal{I}):~~~&\left[d_t-\max\left\{\pi v(\bm d^t),\max_{k\in[t-1]}(d_k-\delta_k)\right\}\right]^++\max_{v_i,\delta_{ij}, d_i,i\in\mathcal{I}}~~\sum_{i\in\mathcal{I}}(d_i-\pi v_i)\\
%       \text{subject to}~&~~~\sum_{j=1}^{T}\delta_{ij}=c,\text{, for all }i\in \mathcal{I};~0\leq \delta_{ij}\leq \bar{\delta}\text{, for all }i\in \mathcal{I}\text{ and }j\in [T];\\
%       &~~~d_j-\delta_{ij}\leq v_i\text{, for all }i\in \mathcal{I}\text{ and } j\leq i;~\underline{d}-\delta_{ij}\leq v_i\text{, for all }t\leq i< j\leq T;\\
%       &~~~\max_{k\in[t-1]}(d_k-\delta_k)\leq d_i\leq \overline{d},\,\max_{k\in[t-1]}(d_k-\delta_k)\leq \pi v_i\text{, for all }i\in \mathcal{I}.
%     \end{split}
%   \end{equation*}

    \begin{equation*}
    \begin{split}
    \textsf{AdaCR-Threshold}&(\pi,\mathcal{I}): [d_t -\max_{k\in[t]}d_k+\sigma(\bm d^t)/\pi]^+\\
& +\max_{v_i,\delta_{ij}, d_i,i\in\mathcal{I}}~~\sum_{i\in\mathcal{I}}\left[d_i -m
_i+\left(m_i-v_i\right)/\pi\right]\\
      \text{subject to}~&\sum_{j=1}^{T}\delta_{ij}=c,\text{ for all }i\in \mathcal{I};\\
      & 0\leq \delta_{ij}\leq \bar{\delta}\text{, for all }i\in \mathcal{I}\text{ and }j\in [T];\\
      & d_j-\delta_{ij}\leq v_i\text{, for all }i\in \mathcal{I}\text{ and } j\leq i;\\
      & \underline{d}-\delta_{ij}\leq v_i\text{, for all }t\leq i< j\leq T;\\
      & d_k\leq m_i\text{, for all }k\in[i] \text{ and }i\in \mathcal{I};\\
      & \underline{d}\leq d_i\leq\overline{d}\text{, for all }i\in \mathcal{I}.
      %&~~~\max_{k\in[t-1]}(d_k-\delta_k)\leq d_i\leq \overline{d},\,\max_{k\in[t-1]}(d_k-\delta_k)\leq \pi v_i\text{, for all }i\in \mathcal{I}.
    \end{split}
  \end{equation*}

The objective function corresponds to the sum of discharging quantities over a set of time slots assuming that we maintain the online-to-offline ratio of peak-demand reduction to be~$\pi$ From current time slot to $T$. Similar to~\textsf{CR-Comp}($\mathcal{I}$), the constraints of~$\textsf{AdaCR-Threshold}(\pi,\mathcal{I})$ are due to the offline {\proname} problem solved in each time slot under the adaptive~\algname.
%while the difference lies in that we never decrease the purchased demand of a future time slot to be less than the observed peak purchased demand, described by the inequality constraints on the last line.
By similar arguments for computing the best CR~$\pi^*$, we conclude that the adaptive CR at time slot~$t$ should be the smallest ratio~$\pi$ in~$\left[\pi_{lb}^t,\pi^*_{t-1}\right]$ such that~$\max_{\mathcal{I}\in \mathcal{I}_t} \textsf{AdaCR-Threshold}(\pi,\mathcal{I})$ does not exceed the remaining inventory~$c-\sum_{k=1}^{t-1}\delta_k$. Therefore, we can search for~$\pi^*_t$ by the bisection method in Algorithm~\ref{algadaptivepCRcompute}. Together with Algorithm~\ref{algadaptivepCRcompute}, we complete the introduction of Algorithm~\ref{algadaptivepCR} and now summarize the theoretical performance in the following proposition.

\begin{proposition}
  At each time slot $t$, Algorithm~\ref{algadaptivepCR} achieves the optimal adaptive competitive ratio among all algorithms in $\mathcal{A}_t$.
\end{proposition}

If the input sequence~$\bm d$ is in the worst case regarding~\algname($\pi^*$), then~\mbox{$\pi^*_t=\pi^*$}, for all~\mbox{$t\in[T]$}. Otherwise, there is an index~$\tau\in[T]$ such that~$\pi^*_t< \pi^*$, for all~$t\geq \tau$. From this perspective, we show that the adaptive~\algname~also attains the optimal CR among all online algorithms for~{\proname}; moreover, it outperforms~\algname($\pi^*$) under general cases. %As a whole, extending~\algname($\pi^*$) to the adaptive~\algname~algorithm can improve the practical utilization of storage for maximizing peak-demand reduction, which will be further verified by real-world traces later. %Such an extension also clarifies the strength of the way we design competitive online algorithms.

\section{Simulation}\label{Sec_Simulation}
In this section, we apply the real-world traces to evaluate the performance of our algorithm under diverse settings. We also compare with conceivable alternatives. The results corroborate our theoretical findings and demonstrate the potentials for practical implementation of our approaches.
\subsection{Simulation Setups}
In the simulation, we consider a scenario where an EV charging station operator uses its storage to reduce its peak demand in a day. We obtain three-month electricity data from an EV charging station in Shenzhen, China. %The data consists of all charging requests of the station in three months. Each request contains its starting time, duration, and charged energy. We first convert the data into the power demand of the charging station at each minute assuming the charged energy is delivered at a constant rate in the charging duration. %As the peak-demand charge is at a length of 15-min,
We divide the time into slots with 15-min length and derive the power demand of the charging station at each slot from the data. We then identify the on-peak duration in a day from the data. In particular, we consider a decision period of $T=20$ slots and set the demand bounds as $(\underline{d},\bar{d})=(442.91,1020.10)$ kWh, which are the minimum and maximum demand of the charging station in this three months, respectively.

\subsection{Performance Evaluation}

\begin{figure}[t]
    \centerline{\includegraphics[width=0.85\columnwidth]{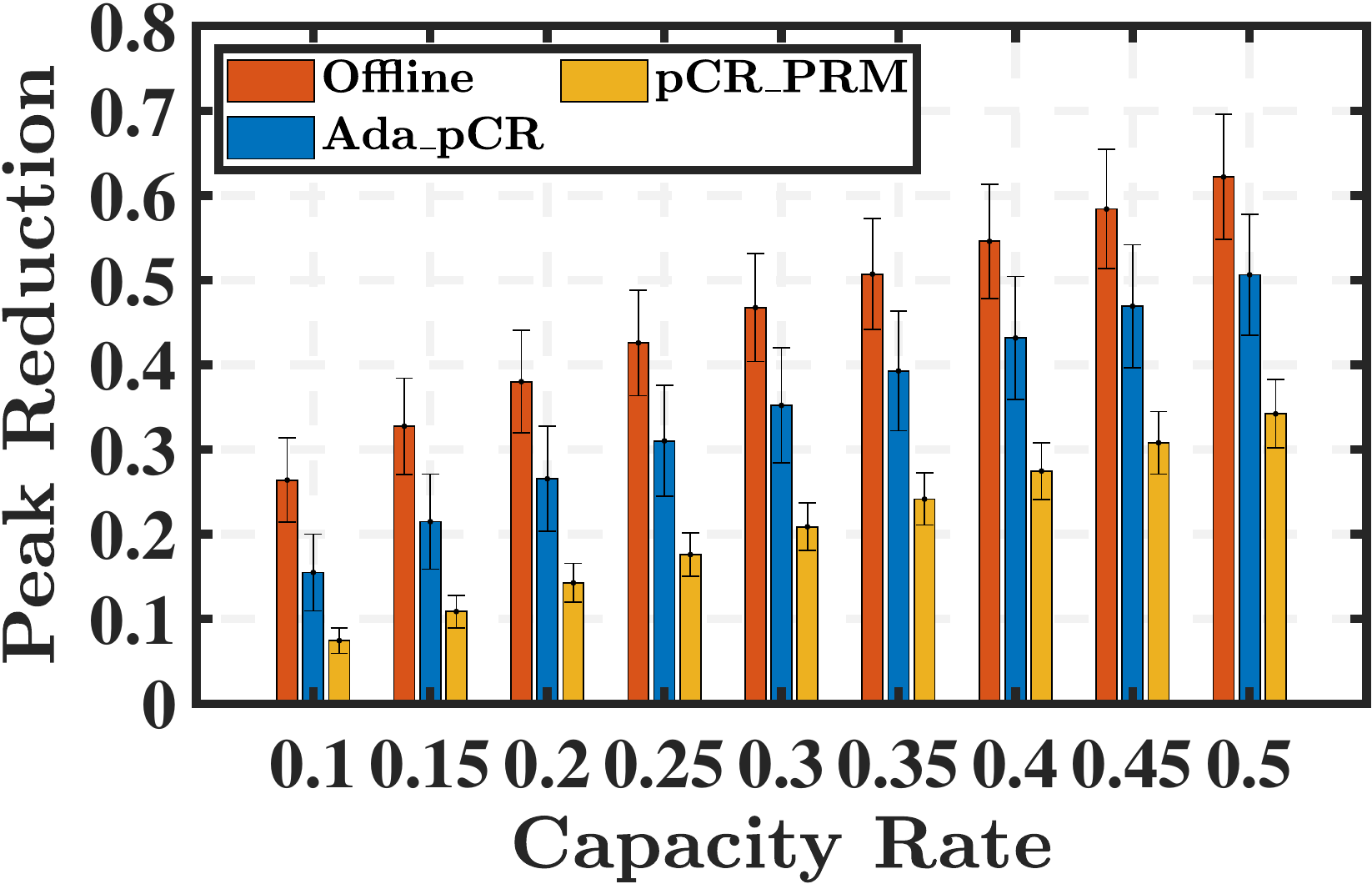}}
    %\captionsetup{format=hang,indention=0.0em, font=footnotesize}%format=hang,
    \caption{Peak reduction under optimal offline solutions, \textsf{pCR-PMD}($\pi^*$), and the anytime-optimal \textsf{pCR-PMD}.}
  \label{fig: comparisons_ada}
  \end{figure}

\textbf{$\algname(\pi^*)$ and adaptive $\algname$} We evaluate the performance of these two algorithms under different storage capacities and show the result in Fig.~\ref{fig: comparisons_ada}. The capacity rate represents the ratio between the storage capacity and the average daily demand of the charging station in the specified on-peak period. The peak reduction refers to the average peak reduction rate, which is the average ratio between the peak reduction achieved by respective algorithms and the original peak demand. For Fig.~\ref{fig: comparisons_ada}, we observe that all the algorithms perform better as the capacity increases. The adaptive $\algname$ attains a higher peak reduction as compared with $\algname$, which corroborates our theoretical findings in Sec.~\ref{Sec_Adaptive}. While with little feature demand information, our online algorithm adaptive $\algname$ can achieve $59\%\sim 81\%$ of the peak reduction of the offline optimal. %When compared with $AO\_PAM$ in ???, adaptive has slight improvements and much smaller deviations on peak reduction under all capacity rates.

\begin{figure}[t]
  \centering

  \begin{subfigure}[t]{\columnwidth}
    \centerline{\includegraphics[width=.85\columnwidth]{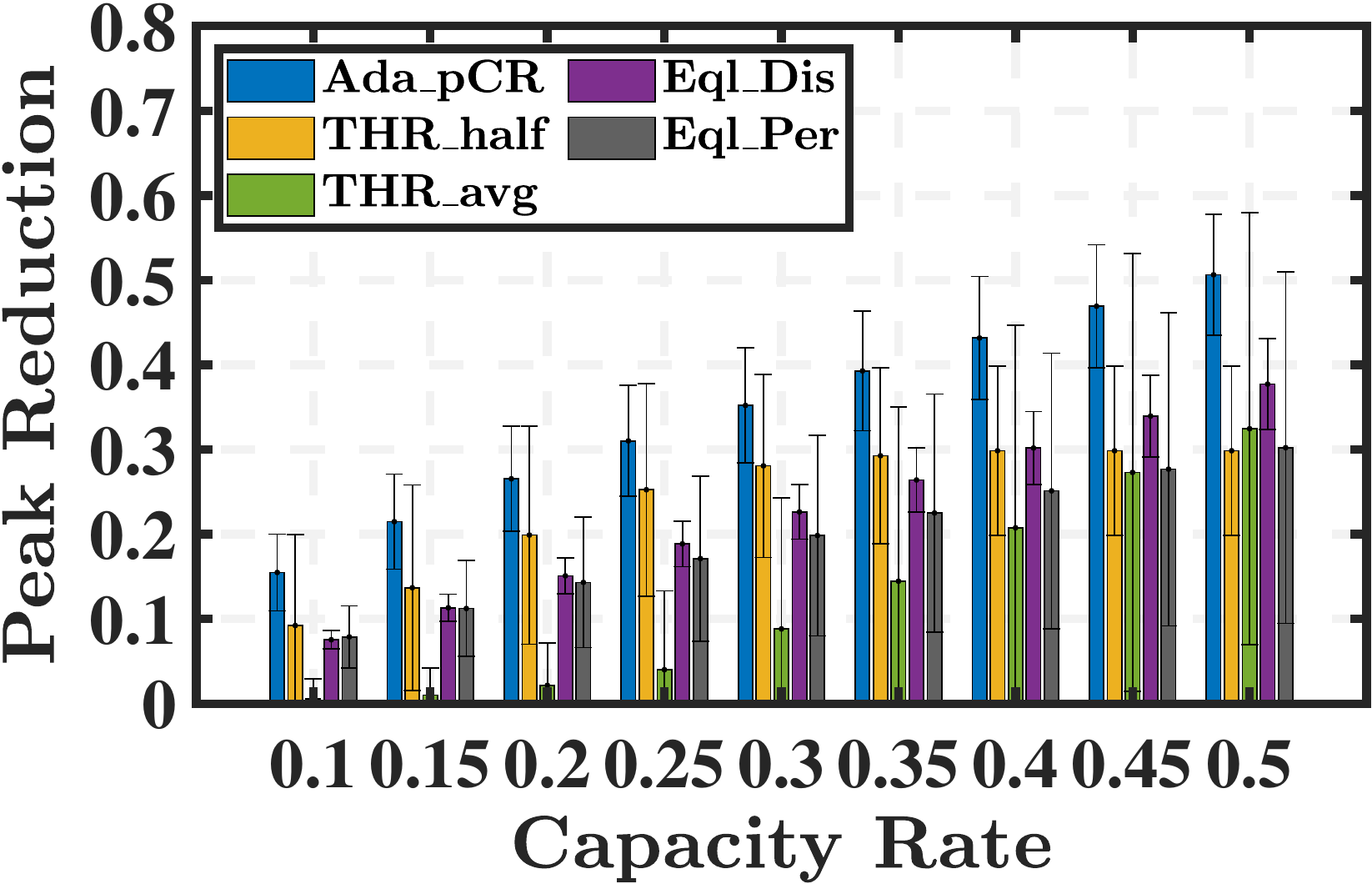}}
    %\captionsetup{format=hang,indention=0.0em, font=footnotesize}
    \caption{} \label{fig: comparisons_alter}
  \end{subfigure}
 \vskip\baselineskip
  \begin{subfigure}[t]{\columnwidth}
    \centerline{\includegraphics[width=.85\columnwidth]{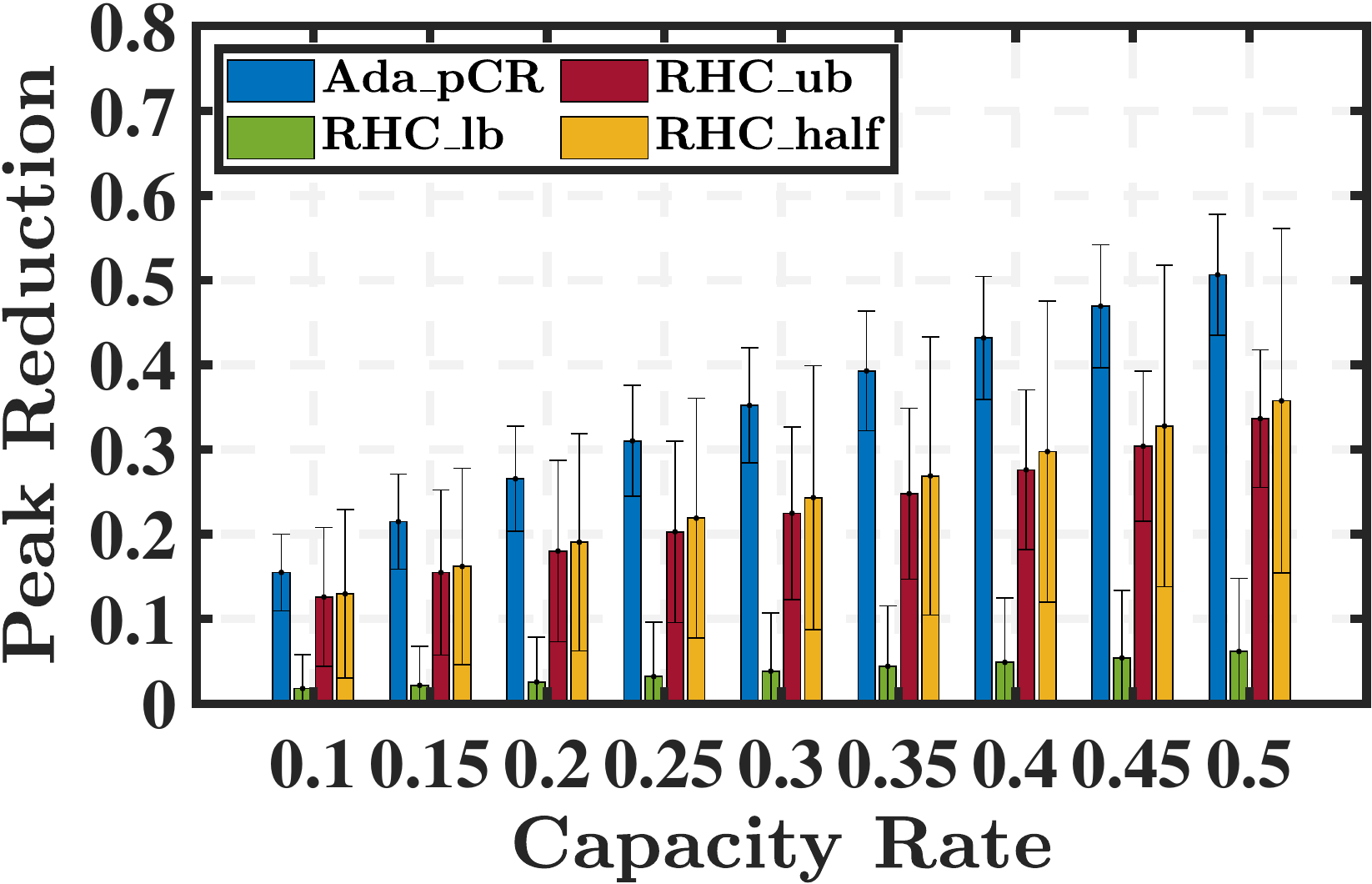}}
    %\captionsetup{format=hang,indention=0.0em, font=footnotesize}
    \caption{}\label{fig: comparisons_rhc}
  \end{subfigure}
  %\label{fig: Sensitivity}
  \caption{Empirical performance comparison as the storage capacity changes. a)~Peak reduction under four threshold-based algorithms and the anytime-optimal \textsf{pCR-PMD}. b)~Peak reduction under three RHC algorithms and the anytime-optimal \textsf{pCR-PMD}.}
\end{figure}

\textbf{Comparison with Alternatives} We mainly compare adaptive $\algname$ with two catalogs of conceivable alternatives, naive threshold-based algorithms and receding horizon control (RHC) algorithms. We show the results in Fig.~\ref{fig: comparisons_alter} and Fig.~\ref{fig: comparisons_rhc}. In particular, we introduce the alternatives as follows,
\begin{enumerate}
    \item  \textsf{THR\_half} and \textsf{THR\_avg} represent the algorithms that discharging to a threshold at each slot until using up the storage. \textsf{THR\_half} sets the threshold as $\left(\underline{d}+\bar{d}\right)/2$, and \textsf{THR\_avg} sets the threshold as the average offline optimal peak demand after using the storage.
    \item \textsf{Eql\_Dis} equally distributes the energy storage capacity at each slot. \textsf{Eql\_Per} discharges at a constant ratio of the demand at each slot until the capacity is running out, and sets the ratio the same as the capacity rate.
    \item RHC algorithms assume a look-ahead window of $5$ slots (a quarter of the on-peak duration) in our simulation. At each slot, the RHC algorithms first compute the optimal solution based on the demand in the look-ahead window and their guesses beyond the window. Then, it implements the optimal solution at the current slot and recomputes the optimal at the next slot. \textsf{RHC\_lb, RHC\_ub, and RHC\_half} assume the demand beyond the look-ahead window as $\underline{d}$, $\overline{d}$, and $\left(\underline{d}+\overline{d}\right)/2$, respectively.
\end{enumerate}
From Fig.~\ref{fig: comparisons_alter}, we observe that adaptive $\algname$ achieves the largest peak-shaving under different capacity rates, with at least~$23\%$ improvement against threshold-based algorithms. We observe from the Fig.~\ref{fig: comparisons_rhc} that adaptive $\algname$ has at least $20\%$ improvement and  a relatively small deviation on the peak reduction as compared to the RHC algorithms.

\begin{figure}[t]
    \centerline{\includegraphics[width=0.85\columnwidth]{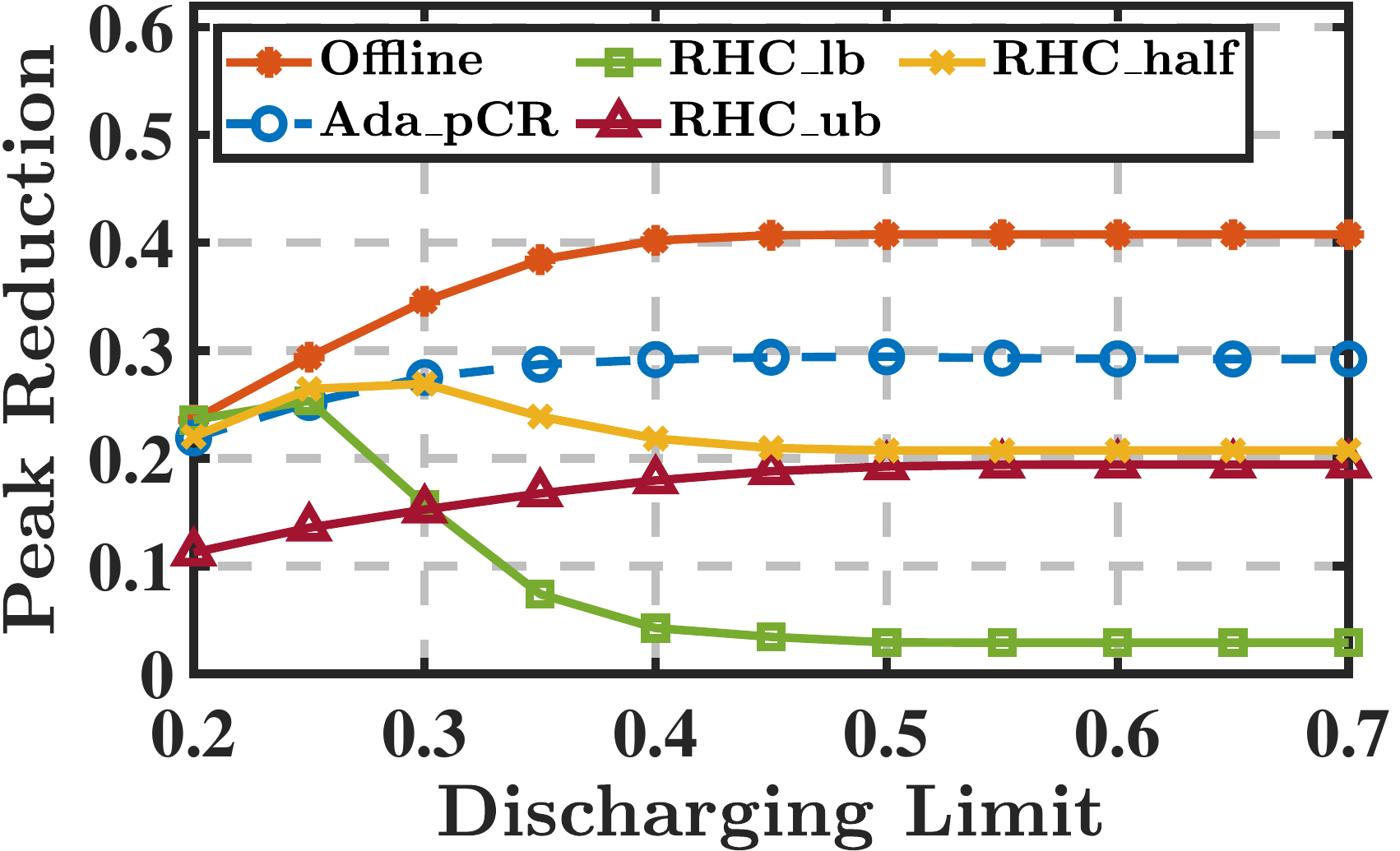}}
    %\captionsetup{format=hang,indention=0.0em, font=footnotesize}
    \caption{Effects of the maximum discharging rate.}\label{fig: comparisons_rate}
\end{figure}

\textbf{Impact of Discharging Limit} We evaluate the performance of adaptive $\algname$ under different maximum charging rates and compare it with alternatives. The results are shown in Fig.~\ref{fig: comparisons_rate}. We normalize the discharging rate limit by the upper bound $\bar{d}$ of the demand at a slot. We observe that the adaptive $\algname$ has close performance under different discharging limits, while the offline optimal achieves greater peak reduction as the discharging limit increases. Without knowing the future demand, adaptive $\algname$ with a larger charging rate limit may waste more energy at the previously observed peak demand which turns out not to be the peak in the whole period. This prevents adaptive $\algname$ to obtain higher peak reduction at a larger discharging rate limit. Besides, adaptive $\algname$ outperforms other alternatives under different charging rate limits.

\section{Conclusion}\label{Sec_Conclusion}

We study an online peak-demand reduction maximization problem under inventory constraints. We focus on a scenario that a large-load power consumer applies energy storage to reduce its peak demand during the on-peak period in a day. We derive an optimal algorithm $\algname(\pi^*)$ that achieves the best CR among all online algorithms. We obtain the optimal CR $\pi^*$ by solving a series of $T$ linear-fractional programs. We further extend our algorithm to an adaptive one by exploiting the observed input in real-time. The adaptive $\algname$ achieves the best adaptive CR at each time slot given the revealed input and online decisions so far. It improves the average-case performance while maintaining the worst-case guarantee. Finally, We evaluate the empirical performance of our algorithms applying real-world traces. We show that the adaptive $\algname$ achieves close performance to the offline optimal and outperforms the conceivable alternatives. \rev{In the future, we shall consider energy storage systems with recharging, e.g., batteries. It is also interesting to extend our techniques for scheduling flexible loads, as in~\cite{zhao2017robust}.}% the charging and discharging simultaneously

\begin{acks}
The work presented in this paper was supported in part by a Start-up Grant (Project No.~$9380118$) from City University of Hong Kong. Qiulin Lin was with the department of Information Engineering, The Chinese University of Hong Kong, during this work.
\end{acks}

%\section{Appendices}
%
%If your work needs an appendix, add it before the
%``\verb|\end{document}|'' command at the conclusion of your source
%document.
%
%Start the appendix with the ``\verb|appendix|'' command:
%\begin{verbatim}
%  \appendix
%\end{verbatim}
%and note that in the appendix, sections are lettered, not
%numbered. This document has two appendices, demonstrating the section
%and subsection identification method.

%\begin{acks}
%To Robert, for the bagels and explaining CMYK and color spaces.
%\end{acks}

%\bibliographystyle{ACM-Reference-Format}

%\newpage
%\begin{APPENDIX}{Proofs}\label{app_proof}
%\section{Notation}\label{app_notation}

\appendix

\section{Proof of Proposition~\ref{offopt}}\label{app_offopt}
We rewrite the optimal-solution formula as follows:
$$\delta_t^*=\begin{cases}
   [d_t-[v]^+]^+ &\text{if } v\geq \max_{t\in[T]}d_t-\bar{\delta};\\
   [d_t-[\max_{t\in[T]}d_t-\bar{\delta}]^+]^+     &\text{otherwise}.
  \end{cases}$$
The existence of~$v$ is trivial and the function~$f(x)=\sum_{t=1}^{T}[d_t-x]$ is nonincreasing in~$v$. Let~$\bm x$ be an optimal solution to \proname~satisfying~$\sum_{t=1}^{T}x_t\leq c$. We first assume that~$v\geq \max_{t\in[T]}d_t-\bar{\delta}$. If~$v\geq 0$, then it follows from~$\sum_{t=1}^{T}[d_t-v]^+=c$ that~$v=\max_{t\in[T]}\left(d_t-x_t\right)$ and~$\delta_t=[d_t-v]^+$ generates a feasible solution attaining the optimum. Otherwise, if~$v<0$, we have~$\sum_{t=1}^{T}d_t<c$ and~\mbox{$d_t\leq \bar{\delta}$}, implying~$\max_{t\in[T]}\left(d_t-x_t\right)=0$. Second, we assume that~$v<\max_{t\in[T]}d_t-\bar{\delta}$. If~$\max_{t\in[T]}d_t> \bar{\delta}$, then~$\max_{t\in[T]}\left(d_t-x_t\right)=\max_{t\in[T]}d_t-\bar{\delta}>0$ and~$\delta_t=[d_t-(\max_{t\in[T]}d_t-\bar{\delta})]^+$ generates a feasible solution attaining the optimum. Otherwise, we have~$\max_{t\in[T]}\left(d_t-x_t\right)=0$ and~$x_t=d_t$, for all~$t\in[T]$. To summarize, we conclude that an optimal solution is given by the formula in Proposition~\ref{offopt}.

\section{Proof of Proposition~\ref{prop_rand}}\label{app_rand}
  Let~$\mathfrak{A}$ be a randomized online algorithm for {\proname}. Without loss of generality, we assume that~$\mathfrak{A}$ is a probability distribution~$\{\omega(\mathfrak{C})\}$ over a set of deterministic online algorithms for {\proname}:~$\mathfrak{C}\in\mathcal{C}$. Then, the expected peak reduction under Algorithm~$\mathfrak{A}$ and a demand profile~$\bm d$ is given by~$$\mathbf{E}[\sigma^\mathfrak{A}(\bm d)]=\int_\mathcal{C}\left(\max_{t\in[T]}d_t-\max_{t\in[T]} (d_t-\delta_t^\mathfrak{C})\right)d\omega(\mathfrak{C}),$$ where~$\delta_t^\mathfrak{C}$ is the discharging quantity at time slot~$t$ under Algorithm~$\mathfrak{C}\in \mathcal{C}$ and the demand profile~$\bm d$.

  We devise another deterministic algorithm~$\mathfrak{B}$ based on the distribution~$\{\omega(\mathfrak{C})\}$ over the algorithm set~$\mathcal{C}$. Specifically, the discharging quantity of time slot~$t$ under Algorithm~$\mathfrak{B}$ and the demand profile~$\bm d$ is given by~$\int_{\mathcal{C}}\delta_t^\mathfrak{C}d\omega(\mathfrak{C})$, for all~$t\in [T]$.
  Since \proname~has linear constraints, it follows that the solution generated by Algorithm~$\mathfrak{B}$ is always feasible for all~$\bm d\in\mathcal{D}$. Moreover, the peak reduction under Algorithm~$\mathfrak{B}$ and the input sequence~$\bm d$ is given by
  \begin{equation*}
  \begin{split}
      \sigma^\mathfrak{B}(\bm d)&=\max_{t\in[T]}d_t-\max_{t\in T} \left(d_t-\int_{\mathcal{C}}\delta_t^\mathfrak{C}d\omega(\mathfrak{C})\right)\\
      &= \max_{t\in[T]}d_t-\max_{t\in T} \left(\int_{\mathcal{C}}\left(d_t-\delta_t^\mathfrak{C}\right)d\omega(\mathfrak{C})\right)\\
      &\geq \max_{t\in[T]}d_t-\int_{\mathcal{C}}\max_{t\in[T]}\left(d_t-\delta_t^\mathfrak{C}\right)d\omega(\mathfrak{C})=\sigma^\mathfrak{A}(\bm d),
  \end{split}
  \end{equation*}
  where the inequality is due to the Jensen's inequality. Thus, we complete the proof.

\section{Proof of Lemma~\ref{lemmaintain}}\label{app_maintain}
It follows from the formula in Algorithm~\ref{algpCS} that $\text{, for all }t\in[T],$
$$d_t-\delta_t\leq \max_{k\in[t]}d_k - \sigma(\bm d^t)/\pi=(1-1/\pi)\max_{k\in[t]}d_k+v(\bm d^t)/\pi.$$
Since~$v(\bm d^t)$ and~$\max_{k\in[t]}d_k$ are non-decreasing in~$t$, we conclude that $\text{, for all } i\in[t],$
$$d_i-\delta_i\leq (1-1/\pi)\max_{k\in[i]}d_k+v(\bm d^i)/\pi\leq (1-1/\pi)\max_{k\in[t]}d_k+v(\bm d^t)/\pi.$$
If follows that~$$\max_{k\in[t]}d_k-(d_i-\delta_i)\geq \max_{k\in[t]}d_k/\pi-v(\bm d^t)/\pi=\sigma(\bm d^t)/\pi\text{, for all } i\in[t].$$ Then, we complete the proof.

\section{Proof of Proposition~\ref{propfeasibility}}\label{app_feasibility}
  The necessary part is obvious and it remains to show the sufficient part. First of all, we show that
  \begin{align*}
      \delta_t(\pi,\bm d) &=[d_t -\max_{k\in[t]}d_k+\sigma(\bm d^t)/\pi]^+\\
    &  \leq  d_t -\left(\max_{k\in[t]}d_k-\sigma(\bm d^t)/\pi\right)
   \leq d_t,
  \end{align*}
  where the second inequality follows from~$\max_{k\in[t]}d_k\geq \sigma(\bm d^t)$ and~$\pi\geq 1$. Then, we show that each solution generated by~\algname($\pi$) satisfies the discharging limit constraint. Specifically, we have
  \begin{align*}
    \delta_t(\pi,\bm d)\leq  d_t -\max_{k\in[t]}d_k+\sigma(\bm d^t)/\pi \leq  \sigma(\bm d^t)/\pi\leq \sigma(\bm d^t) \leq \bar{\delta},
  \end{align*}
   where the last inequality is due to the feasibility of an offline optimal solution to {\proname}. Thus, we complete the proof by observing that~$\bm \delta(\pi,\bm d)$ is a feasible solution if and only if it satisfies the inventory constraint.

\section{Proof of Lemma~\ref{lemdeinc}}\label{app_deinc}
  Since~$\delta_t(\pi,\bm d)$ is nonincreasing in~$\pi$, we conclude that~$\Phi(\pi)$ is nonincreasing. Moreover, if~$\Phi(\pi)>0$, then there exist~$\bm d\in\mathcal{D}$ and~$k\in[T]$ such that~$\Phi(\pi)=\sum_{t=1}^{T}\delta(\pi,\bm d)$ and~$\delta_k(\pi,\bm d)>0$. Note that~$\delta_k(\pi,\bm d)$ is strictly decreasing when~$\delta_k(\pi,\bm d)>0$. It follows that~$\Phi(\pi)$ is strictly decreasing when~$\Phi(\pi)>0$.

\section{Proof of Theorem~\ref{thmoptratio}}\label{app_optratio}

  It follows from Lemma~\ref{lemdeinc} that the solution to~$\Phi(\pi)=c$ exists and is unique. Moreover, by Proposition~\ref{prop_rand}, we can prove the theorem by showing that~\algname($\pi^*$) is better than any other deterministic online algorithm~$\mathfrak{A}$ for {\proname}. We denote by~${\bm \delta}(\mathfrak{A},\bm d)$ the output sequence under Algorithm~$\mathfrak{A}$ and the input sequence~$\bm d$. Since~$\Phi(\pi^*)=c$, there exists an input sequence~$\hat{\bm d}$ such that~$\sum_{t=1}^{T}\delta_t(\pi^*,\hat{\bm d})=c$. Next, we shall show that Algorithm~$\mathfrak{A}$ cannot attain a strictly smaller CR than~\algname($\pi^*$).

  First of all, we set~$d_1=\hat{d}_1$. If~${\delta}_1(\mathfrak{A},\bm d)<\delta_1(\pi^*,\bm d)$, then we set future inputs by~$d_t=\underline{d}$, for all~$t>1$, and show that
  \begin{align*}
   & \max_{t\in[T]} d_t-\max_{t\in[T]} (d_t-{\delta}_t(\mathfrak{A},\bm d))\\
    \leq &\max_{t\in[T]} d_t-(d_1-{\delta}_1(\mathfrak{A},\bm d))\\
   < &\max_{t\in[T]} d_t-(d_1-{\delta}_1(\pi^*,\bm d))= \sigma(\bm d)/\pi^*,
  \end{align*}
  where the last equality is due to the facts that~$d_1=\max_{t\in[T]}d_t$ and~${\delta}_1(\pi^*,\bm d)=\sigma(\bm d)/\pi^*$. Thus, we conclude that~${\delta}_1(\mathfrak{A},\bm d)\geq \delta_1(\pi^*,\bm d)$, for any algorithm~$\mathcal{A}$ which attains a CR no smaller than~$\pi^*$.

  If~${\delta}_1(\mathfrak{A},\bm d)\geq \delta_1(\pi^*,\bm d)$, then we continue by setting~$d_t=\hat{d}_t$. If~$\bm \delta(\pi^*,\bm d)={\bm \delta}(\mathfrak{A},\bm d)$, then we find the largest time index~$\hat{t}$ such that~$\delta_t(\pi^*,\bm d)>0$. Such an index exists, since~$\sum_{t=1}^{T}\delta_t(\pi^*,\hat{\bm d})=c$. Then, under the input sequence~$\bm d=[d_1~d_2~\cdots~d_{\hat{t}}~\underline{d}~\cdots,\underline{d}]$, the offline-to-online ratio of peak reduction under~$\mathfrak{A}$ remains~$\pi^*$. As a result, the CR~$\mathfrak{A}$ should be no less than~$\pi^*$. If~$\bm \delta(\pi^*,\bm d)\neq {\bm \delta}(\mathfrak{A},\bm d)$, then it follows from~$\sum_{t=1}^{T}\delta_t(\pi^*,\hat{\bm d})=c$ and~$\sum_{t=1}^{T}\delta_t(\mathfrak{A},\hat{\bm d})\leq c$ that~${\delta}_{\hat{t}}(\mathfrak{A},\bm d)< \delta_{\hat{t}}(\pi^*,\bm d)$ for a certain index~$\hat{t}>1$. Then, under the input sequence~$\bm d=[d_1~d_2~\cdots~d_{\hat{t}}~\underline{d}~\cdots,\underline{d}]$, we have
  \begin{align*}
    &\max_{t\in[T]} d_t- \max_{t\in[T]} (d_t-{\delta}_t(\mathfrak{A},\bm d))\\
    \leq & \max_{t\in[T]} d_t-\left(d_{\hat{t}}-{\delta}_{\hat{t}}(\mathfrak{A},\bm d)\right)\\
    < & \max_{t\in[T]} d_t-\left(d_{\hat{t}}-{\delta}_{\hat{t}}(\pi^*,\bm d)\right)=\sigma(\bm d)/\pi^*,
  \end{align*}
  where the last equality follows from the formula of~\algname($\pi^*$). Thus, we conclude that the CR of Algorithm~$\mathfrak{A}$ is strictly larger than~$\pi^*$.

  By the above analysis, we conclude that no other deterministic online algorithm attains a CR strictly smaller than~$\pi^*$. Moreover, we see that the CR of~\textsf{pCR-PAD}($\pi^*$) is~$\pi^*$. Therefore, we complete the proof by further applying Proposition~\ref{prop_rand}.

\section{Proof of Lemma~\ref{lemparwc}}\label{app_parwc}
The existence of a worst-case input sequence for~\algname($\pi^*$) follows from Theorem~\ref{thmoptratio}. Considering a worst-case input sequence~$\hat{\bm d}$, we assume that there is an index~$i
\in[T]$ such that
\begin{equation*}
  \hat{d}_i -\max_{k\in[i]}\hat{d}_k+\sigma(\bm \hat{d}^i)/\pi\leq 0 \text{ and } \hat{d}_{i+1} -\max_{k\in[i+1]}\hat{d}_k+\sigma(\bm \hat{d}^{i+1})/\pi>0.
\end{equation*}
By Proposition~\ref{offopt}, we have~$\max_{k\in[i]}\hat{d}_k-\sigma(\bm \hat{d}^i)/\pi\leq \max_{k\in[i+1]}\hat{d}_k-\sigma(\bm \hat{d}^{i+1})/\pi$. It follows that~$\hat{d}_i< \hat{d}_{i+1}$.

Then, we exchange the order of~$\hat{d}_i$ and~$\hat{d}_{i+1}$ in~$\hat{\bm d}$ and obtain a new input sequence. We show that
\begin{align*}
  \tilde{d}_i -\max_{k\in[i]}\tilde{d}_k+\sigma(\bm \tilde{d}^i)/\pi \geq \hat{d}_{i+1}-\max_{k\in[i+1]}\hat{d}_k+\sigma(\bm \hat{d}^{i+1})/\pi>0 \text{ and }\\
  \tilde{d}_{i+1} -\max_{k\in[i+1]}\tilde{d}_k+\sigma(\bm \tilde{d}^{i+1})/\pi \leq \hat{d}_{i}-\max_{k\in[i]}\hat{d}_k+\sigma(\bm \hat{d}^{i})/\pi\leq 0.
\end{align*}
It follows that~$\delta_i(\pi^*,\tilde{\bm d})\geq \delta_{i+1}(\pi^*,\hat{\bm d})$ and~$\delta_i(\pi^*,\hat{\bm d})=\delta_{i+1}(\pi^*,\tilde{\bm d})=0$. For any index~$t\in[T]$ other than~$i$ and~$i+1$, we also observe that~$\delta_t(\pi^*,\tilde{\bm d})=\delta_t(\pi^*,\hat{\bm d})$. We summarize the discharging quantities over the~$T$ time slots and obtain that
  \begin{equation*}
    c=\Phi(\pi^*)\geq \sum_{t=1}^{T}\delta_t(\pi^*,\tilde{\bm d})\geq \sum_{t=1}^{T}\delta_t(\pi^*,\hat{\bm d})=c,
  \end{equation*}
It follows that~$\sum_{t=1}^{T}\delta_t(\pi^*,\tilde{\bm d})=c$ and~$\tilde{\bm d }$ is also a worst-case input sequence for~\algname($\pi^*$). By applying a sequence of exchanges mentioned above, we can finally attain a worst-case input sequence as described in the theorem, which completes the proof.

\end{document}